\documentclass[12pt]{article}
\usepackage[OT1]{fontenc}

\RequirePackage{amsmath,amsfonts,amssymb}
\usepackage{multirow}
\usepackage{hhline}
\usepackage{subcaption}
\usepackage[utf8]{inputenc}
\usepackage{url}
\usepackage{graphicx}
\usepackage{times}
\usepackage{adjustbox}
\usepackage{tabularray}
\usepackage{cprotect}
\usepackage{lettrine}
\usepackage{textcomp}
\usepackage[dvipsnames]{xcolor}
\usepackage{lineno, blindtext}

\usepackage[labelsep=period]{caption}
\usepackage[labelfont=bf]{caption}

\title{An experimental study of the influence of anonymous information on social media users}

\author
{
    Boleslaw K. Szymanski$^{1,*,\dagger}$, Brendan Cross$^{1,\dagger}$, 
    John Hulton$^{1}$,\\ 
    James Flamino$^{1}$, Chris Gaiteri$^{2}$, Jonathan Z. Bakdash$^{3}$,\\
    \normalsize{$^{1}$Department of Computer Science and Network Science and Technology Center,}\\
    \normalsize{Rensselaer Polytechnic Institute, Troy, NY, USA}\\
    \normalsize{$^{2}$Department of Psychiatry and Behavioral Sciences,}\\
    \normalsize{SUNY Upstate Medical University, Syracuse, NY, USA;}\\
    \normalsize{$^{3}$Department of Psychology and Special Education, Texas A\&M -- Commerce,}\\
    \normalsize{Commerce, TX, USA}\\
    \normalsize{$^{*}$Corresponding author, Email: szymab@rpi.edu}\\
    \normalsize{$^{\dagger}$These authors contributed equally to this work.}
}

\date{}

\begin{document}

\baselineskip24pt

\maketitle

\begin{abstract}
Increasingly, people use social media for their day-to-day interactions and as a source of information, even though much of this information is practically anonymous. This raises the question: does anonymous information influence its recipients? We conducted an online, two-phase, preregistered experiment using a nationally representative sample of participants from the U.S. to find the answer. To avoid biases of opinions among participants, in the first phase, each participant examines ten Rorschach inkblots and chooses one of four opinions assigned to each inkblot. In the second phase, the participants are randomly assigned to one of four distinct information conditions and are asked to revisit their opinions for the same ten inkblots. Conditions ranged from repeating phase one to receiving anonymous comments about certain opinions. Results were consistent with the preregistration. Importantly, anonymous comments shown in phase two influence up to half of the participants' opinion selections. To better understand the role of anonymous comments in influencing the selections of opinions, we implemented agent-based modeling (ABM). ABM results suggest that a straightforward mechanism can explain the impact of such information. Overall, our results indicate that even anonymous information can have a significant impact on its recipients, potentially altering their popularity rankings. However, the strength of such influence weakens when recipients' confidence in their selections increases. 
Additionally, we found that participants' confidence in the first phase is inversely related to the number of change opinions.

\end{abstract}

\section*{Introduction}

Making choices on political issues, or voting for candidates in elections, is a process which has changed significantly over the last few decades because of the growth of social media. Human interactions developed over centuries of face-to-face interaction practices that limited the number of potential acquaintances and relied on the confidence gained during the conversations and guided by the body language and social status of the interacting parties. In such a setting, two trends have arisen. One is a confirmation bias, which is a tendency to keep already established opinions and beliefs. Another is homophily, which is the desire to hold opinions and beliefs that are compatible with many members of the social groups or societies, and therefore popular \cite{flamino2021creation}. Today, cellphones, online interactions and social media vastly expand the sizes of social groups, and the number of people with whom an ordinary person interacts. In such an environment, the instantaneous reaction of each person to events in their lives and the world must be considered alongside the influence of social commentary. Sharing personal responses in the modern group setting intrinsically generates tension between self-expression versus mirroring group norms.  Self-expression, as shown by individuality or creative behavior, can benefit problem solving \cite{hong2004groups}, while mirroring popular opinions or remaining in large groups offers social rewards \cite{castanon2023reinforcement}, power \cite{pulles2017likeability}, health \cite{bennett2006effect}, and safety benefits \cite{delgado2023characterizing}, with occasional weakness due to reluctance to accept novelty \cite{siler2015measuring,packalen2020nih}. 

Freedom of movements among social media platforms enables people with unpopular opinions, who have difficulty forming in-groups driven by homophily, to either change unpopular opinions to more popular ones \cite{block2014multidimensional} or leave the current environment \cite{flamino2021creation} in a manner dependent on the group around them \cite{muchnik2013social}. Such responses are commonly observed in face-to-face interpersonal interaction groups ranging from close associates \cite{mcpherson1987homophily} to strangers \cite{Levitan_Verhulst_2016}. In the past, the success of changing other's personal opinions correlates with several personal traits of those engaging in influencing others, such as appearance \cite{Laustsen_Petersen_2016,Fernandes_Nettleship_Pinto_2022}, personality \cite{Breves_Liebers_Abt_Kunze_2019},  perceptiveness \cite{zerubavel2015neural} and  opinion domain credibility \cite{Zimmerman_Garbulsky_Ariely_Sigman_Navajas_2022}.  
Currently, confirmation bias and homophily are still active, but without body language clues how much information from the sender can be trusted. This raises the question about the magnitude of influence which information from unknown online sources can exert on social media users. 

To answer this question, we designed a two-phase experiment using a set of Rorschach inkblots, each with four predefined opinions about what the inkblot represents. In the first phase of the study, we gather from each participant their selected opinion, the level of confidence, and a comment explaining their selections of opinion about each inkblot. The second phase separates participants into four treatment groups proceeding under different conditions. We abbreviate condition names as follows: C - confirmation bias, CR - reinforced confirmation bias, CM - most-popular-influence, CS - second-most-popular-influence.

We choose Rorschach inkblots \cite{Rorschach1948} as stimuli because they are ambiguous, and the choice of which opinion best represents an inkblot is a matter of personal preference that is open to reinterpretation. Unlike much prior research utilizing inkblots, we do not seek to assign any specific meaning to inkblots from the responses of individuals, but simply utilize them for their ambiguity itself.

Both the experimental design and the confirmatory analyses were pre-registered at \url{https://aspredicted.org/LGJ_Q3}. Our main pre-registered hypothesis about the second phase was that participants' choices of opinions would change after these participants receive additional information. All participants will more frequently choose the opinion promoted by received comments. We discuss the consequences of this change of behavior in the Results and Discussion sections.

\section*{Experiment Design}
\label{sec:Experiment}
 
The experiment is designed to study how external influences affect participants' choices of opinions about inkblots. But it can also be seen as a high-level abstraction of elections, in which participants are voters and opinions of each inkblots are the candidates' positions on important topics. For each  condition we create a unique way for influencing the participants to change their votes after receiving information sent by the influencers. Designing the experiment that way makes results independent of mundane details of each real election and of existing political biases of participants.

We conducted our experiment on a representative sample of the U.S. population recruited by the Prolific platform, see \url{https://prolific.co}. A total of $N_{p1}=300$ participants completed the first phase and $N_{p2}=242$ returned for the second phase, yielding the completion rate of 80.67\%, exceeding $200$, the pre-registered minimum number of participants.
The opinion analyses include only participants who finished both phases. We summarize the participants' demographic information statistics in the SM Section 1.1. The research received an IRB exemption from Rensselaer Polytechnic Institute, \textbf{Exempt Protocol-IRB ID: 1970, approved on January 28, 2021}.

\begin{figure}[ht!]
	\centering
        \includegraphics[width=1.0\textwidth]{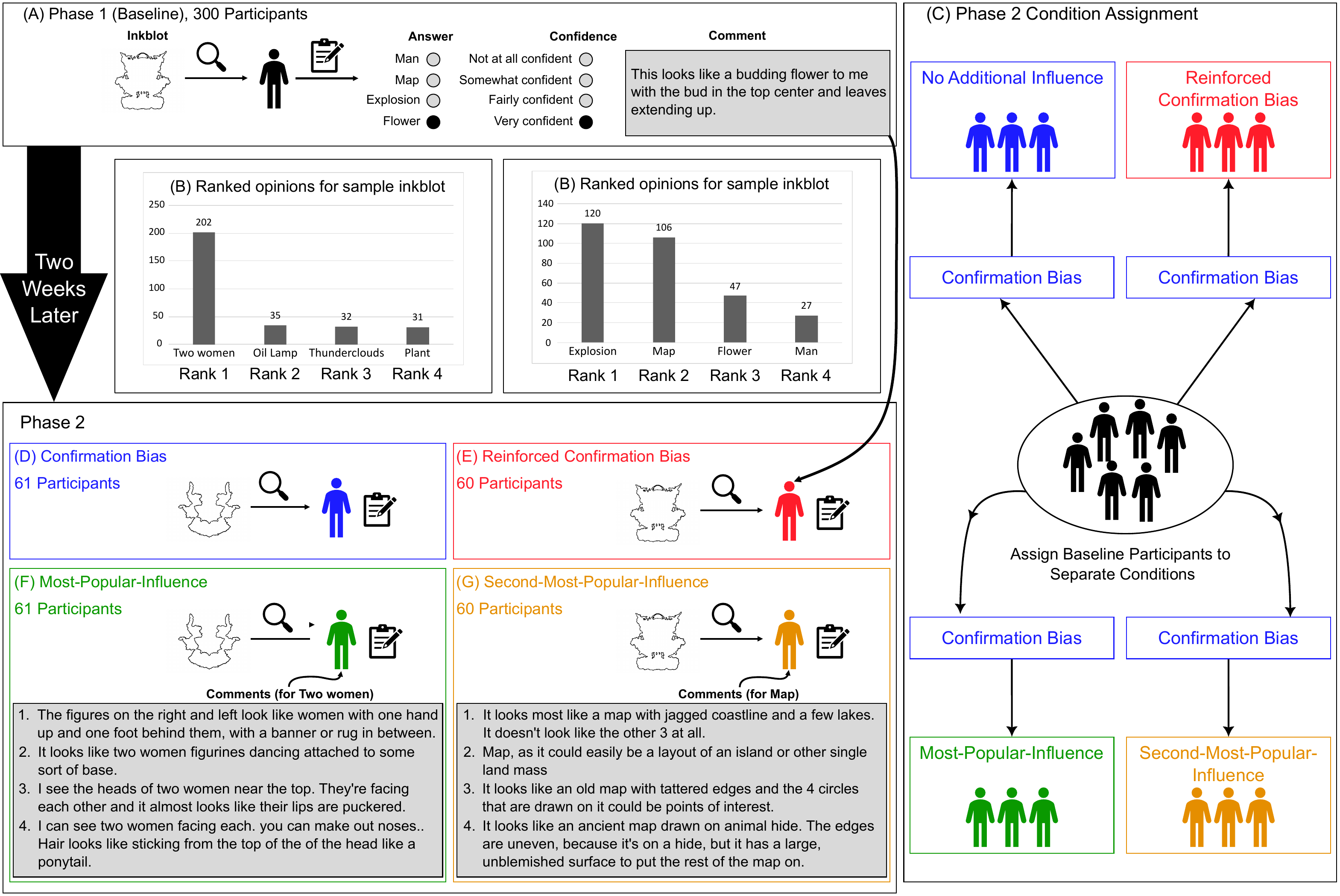}
\caption{{\small{\textbf{Two-phase design of experiment.} The first phase establishes the baseline for all participants. The second measures ambiguous influence under four different conditions.  }}}

    \label{fig:experimental_design}
\end{figure}

Fig. 1 contains an overview of the experiment design. The first phase of the experiment presents all participants with the collection of 10 randomly ordered Rorschach inkblots. Each inkblot has four  opinions describing what it might represent. The participants were required to: (i) choose the opinion best matching their own interpretation of inkblot image, (ii) evaluate their confidence in their choice, and (iii) write a comment explaining their answer, see Fig. 1A. The results of this part of the experiment forms a baseline for frequencies of selections of individual opinions for each inkblot without any influence. We also ranked opinions for every inkblot in decreasing order of their popularity among participants, Fig. 1B. Opinions with rank 1 or 2 are called {\it popular opinions}.

In the second phase, we divide participants into four conditions, Fig. 1C, each presenting participants with the different amount and content of additional information about the opinions. Then participants were asked to process the same collection of inkblots and complete the same tasks as in the first phase, Fig. 1[D-G].  
We use the two-letter abbreviations for conditions in which the confirmation bias influence~\cite{Nickerson1998confirmation}, acquired by selecting opinion in the first phase, is combined with another influence that is active in this condition.

In the SM Section 1.2, we describe how the comments presented to participants in the CM and CS conditions were selected.  The same note about comments is shown to participants in the CM and CS conditions, without information if these comments promote the most-popular or the second-most-popular opinion. No reward or goal was given to participants to sway them to select a particular opinion. No communication among the participants was allowed during and between phases.

The influence of participants in conditions is qualitatively different. The participants in the first two conditions, C and CR, rely on the individual cognition of each participant invoked by their own comments. We refer to such influences as internal. Whereas the participants in the other two conditions receive comments of other participants, and partial information about popularity of the opinion about which these comments were made. Accordingly, we call such influence  external. An opinion about which comment(s) are shown to participants is called the \emph{promoted} opinion.

\section*{Results}
\subsection*{Notation}
The dynamics of the system are captured by each participant's transitions, which are pairs of opinions held by a participant in the first and second phases, denoted as $O^1_r \rightarrow {O^2_r}$, where subscript $r$ denotes the opinion ranks and each superscript denotes the phase. A set of opinions can have several ranks listed in the subscript. All participant holding the promoted opinion for their condition are denoted as $O^p_P$ and referred to as {\it P participants}, while the remaining participants in this condition are denoted as $O^p_N$ and referred to as N participants.

The significantly fewer participants select opinions with the last two ranks than the number of participants selecting the popular opinions. Attributes of the former participants are similarly distributed, making them statistically equivalent to each other (see fig. S8 in SM). Merging rank 4 into rank 3 stabilized the latter size and accelerated convergence of the regression models, outweighing a disadvantage of a small decline in precision of the opinions' frequencies caused by this change.

\subsection*{Analysis of experiment dynamics}

Our main pre-registered hypothesis was that participants' selection of opinions in the second phase would statistically significantly increase the transitions rates to the promoted opinion in each condition. We verify this hypothesis in two ways. 

The first method designed several Bayesian multilevel multinomial regression models implemented using \emph{brms} \cite{burkner2017brms} and \emph{RStan} \cite{stan2020rstan} tools. We fit and compared resulting models with varying numbers of parameters. See table S1 and the details of the selection process are presented in the SM Section 1. Here, we discuss the results of the best performing model that demonstrates how the frequency of opinions in the second phase were affected by the experimental conditions and the rank of the opinion held by a participant. The model predicted frequencies of each opinion transition to promoted opinion are compared to the frequency of the promoted opinion in the first phase, without any influence. 

\begin{figure}
    \centering
        \begin{subfigure}[h]{0.85\textwidth}
        \centering
        \caption{}
        \label{fig:condition_rank_mean_comparisons}
        \includegraphics[width=1\linewidth]{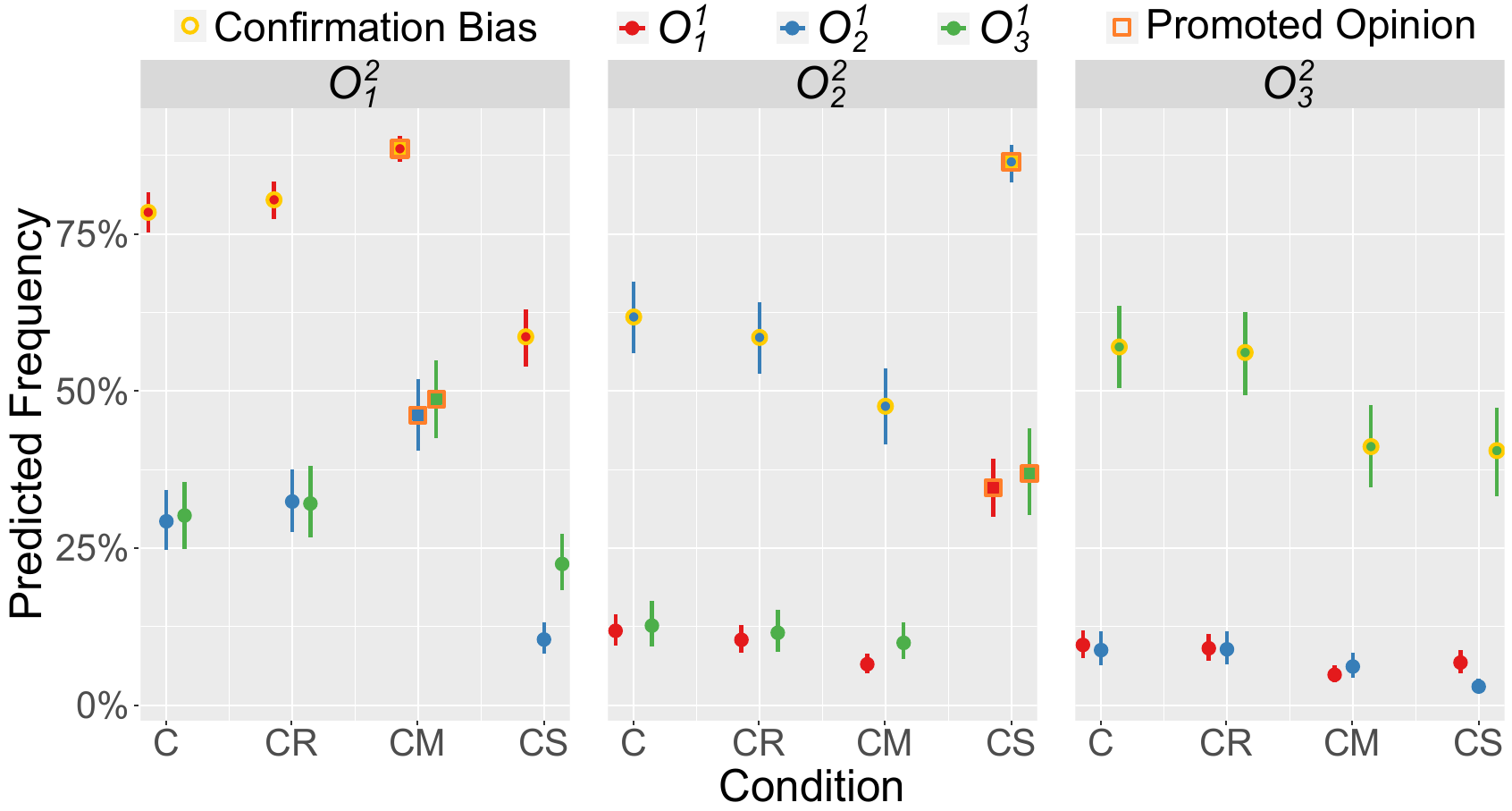}
    \end{subfigure}\\
    \begin{subtable}[h]{0.85\textwidth}
        \centering
        \caption{}
        \label{tab:bayes_regression_table}
        \begin{adjustbox}{width=1\columnwidth,center}
        \begin{tabular}{|rrr|rr|}
          \hline
          & \multicolumn{2}{c|}{$O_1^2$} & \multicolumn{2}{c|}{$O_2^2$} \\
          Predictors & Odds Ratios & CI (95\%) & Odds Ratios & CI (95\%) \\
          \hline
          {Condition C, $O_3^1$ (Intercept)} & 0.60  & 0.32 - 1.10     & 0.29  & 0.14 - 0.57        \\
          Condition CR                  & 1.12  & 0.73 - 1.74     & 0.98  & 0.63 - 1.56        \\
          Condition CM                  & 2.37  & 1.55 - 3.63     & 1.09  & 0.68 - 1.75        \\
          Condition CS                  & 0.98  & 0.62 - 1.55     & 4.34  & 2.76 - 6.87        \\
          $O_1^1$& 14.45 & 10.39 - 20.28   & 3.97  & 2.71 - 5.94        \\
          $O_2^1$& 5.16  & 3.36 - 7.95     & 22.00 & 13.99 - 35.02      \\ 
          \hline
           \multicolumn{5}{|c|}{Random Effects} \\
           \hline
           $\sigma^2_{user}$ & 0.59 & 0.39 - 0.77 & 0.42 & 0.09 - 0.66 \\
           $\sigma^2_{inkblot}$ & 0.73 & 0.41 - 1.35 & 0.85 & 0.47 - 1.59 \\
           \hline
        \end{tabular}
        \begin{tabular}{|l|l|}
             \hline
            	Observations & 2420 \\
                $N_{user}$ & 242 \\
                $N_{inkblot}$ & 10 \\
                ELPD & -1748.8 \\
                ELPD s.e. & 35.8 \\
                LOOIC & 3497.6 \\
                LOIC s.e. & 71.6 \\
                WAIC & 3496.2 \\
        	\hline
        	\end{tabular}
         \end{adjustbox}
    \end{subtable}\\
    \caption{\textbf{Bayesian multi-nomial second phase opinion selection model.} (\subref{fig:condition_rank_mean_comparisons}) highlights the results of the model, where each panel shows the predicted opinion selection frequency for each opinion in the second phase, for each initial opinion and condition combination. (\subref{tab:bayes_regression_table}) lists the parameters and coefficients, presented as odds ratios, of this regression model.}
    \label{fig:bayes_regression_model}
\end{figure}

Fig.~\ref{fig:condition_rank_mean_comparisons} displays the behavior verifying our main hypothesis about the strength of influence exerted on participants for their choices of opinion in the second phase displayed in three panels. Each panel displays the frequencies with which participants chose each promoted opinion averaged over all P participants and each not promoted opinion averaged over all N participants. Predictions for each condition have a unique color blot located at the average frequency and the vertical arms of the same color marking 95\% prediction intervals for participants in this condition.
We surround the blot with the orange square frame for the promoted opinion and with the yellow ring for the frequency of the opinion under confirmation bias, superimposing both when the promoted opinion is also under confirmation bias.

The full definition of the model is given in Fig~\ref{tab:bayes_regression_table}. The dependent variables are the selections of opinions in the second phase. The model accounts for designed influences of conditions and subgroups, and for the random effects caused by each participant having 10 trials to select the second phase opinion, one for each inkblot. The  control condition is the reference group $O_3^{1,2}$ for the first and second phases of opinions of all ranks. 

The model shows a significant increase in selection of the promoted opinions by participants in the CM and CS conditions. When compared to the increase in the C condition, participants in the CM or CS conditions have higher odds of selecting their promoted opinions by the factors of $2.37$ ($95\%$ CI: $1.55 - 3.63$) and $4.34$ ($95\%$ CI: $2.76 - 6.87$), respectively. The same comparison with the increase in CR yields the smaller odds increase. 
In table S2, we outlined a logistic regression model that simplifies this analysis for condition CR. This model predicts the frequency of participants changing opinions from phase 1 to phase 2 per condition, which is the equivalent of not selecting the promoted opinion in phase 2 for the CR condition. This model shows the transition frequency decreases by a factor of $0.63$, the inverse of this shows the increase of promoted opinion selection (mean odds-ratio: $1.63$, $95\%$ CI: $1.09 - 2.56$). The reason CR has a smaller increase in promoted opinion selection is that participants in CM and CS will see five comments selected from the six top quality comments for each promoted opinion. In contrast in CR, each participant will see their single comment from phase one as the promoted opinion. Hence, most of the participants in CR see lower quality comments than the top quality of the five comments shown to participants in CM and CS. The drop in the fraction of participants selecting promoted opinions in CR versus CM and CS implies that low quantity and quality of information spread by influencers weakens their influence.

\subsection*{Tipping points for switching ranks of opinions in phase two}
\begin{figure}
\centering
\includegraphics[width=0.65\textwidth]{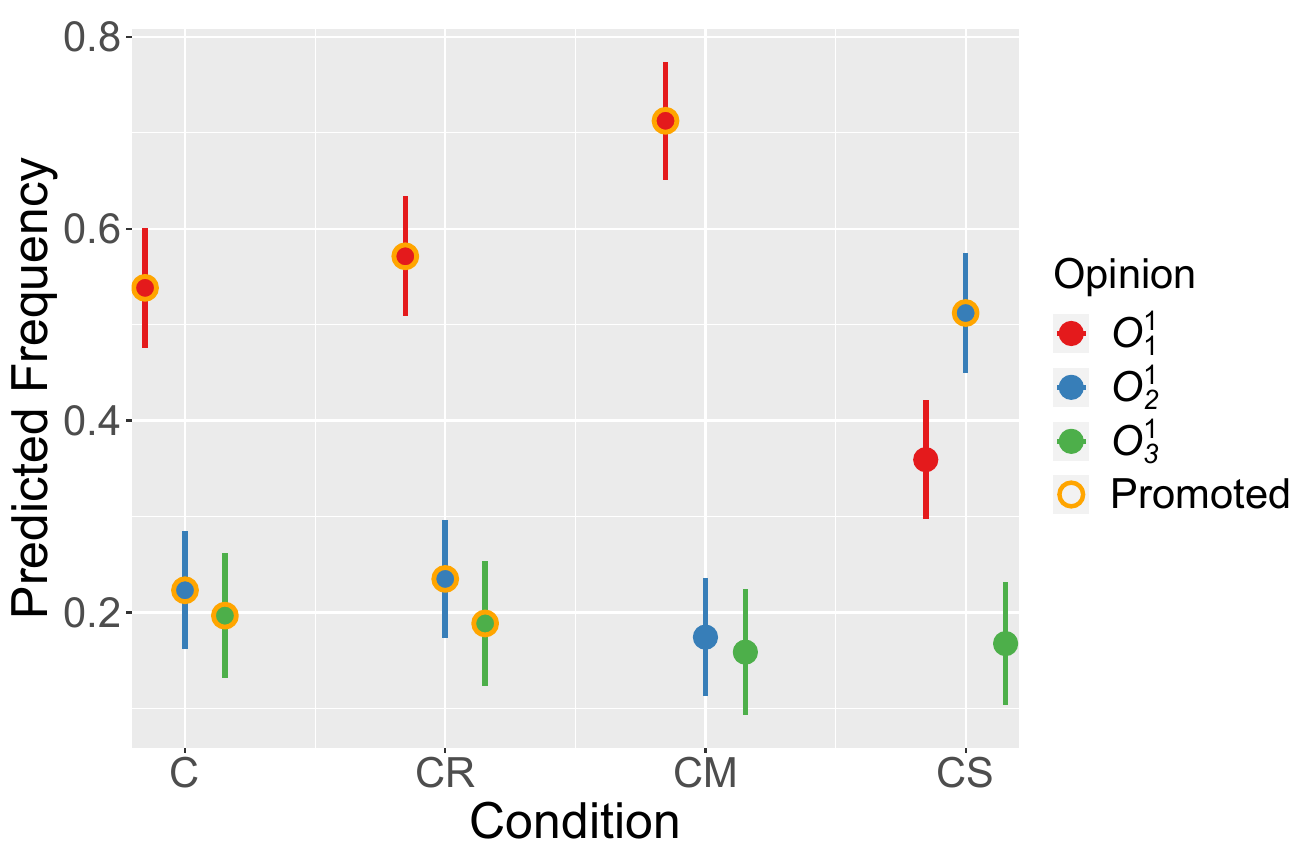}
\caption{\textbf{Predicted final opinion frequency by condition}.
    This figure shows, for each condition, the predicted final selection frequency of each $O_i^2$. }
\label{fig:predicted_final_opinion_frequency}
\end{figure}

Another important behavior seen in our results is the CS condition's promoted comments can change which opinion is the most popular in phase two. Figure \ref{fig:predicted_final_opinion_frequency} shows the predicted selection frequency of each opinion in phase two for each condition. We compute these results using a simple version of the Bayesian multilevel model defined in table \ref{tab:bayes_regression_table}, in which we predict overall selection frequency per phase two opinion. For all conditions, except CS, the $O_i^2$ predicted frequencies have the same order as their $O_i^1$ counterparts. In CS an interesting behavior
arises when the influencing comments increase the selection of the $O_2^2$ to the highest selection frequency. In most cases, the  external influence in the CS condition is strong enough to make the formerly most popular $O_1^1$ opinion less popular than $O_2^1$ in the second phase. This reveals that ambiguous comments can have a fundamental impact on the result of voting in elections.

We next examine how often opinions increase their popularity rank between the first and second phase, $O_i^1 \to \hat{O_j^2}$ where $i < j$, in our dataset. In our experiment, $45\%$ of opinions of all inkblots are switching their ranks in the second phase, and $70\%$ of opinions of 10 inkblots presented in CS condition switched their top two ranks. We found that rank switching of the two most popular opinions in CS condition can be predicted by two simple and intuitive features. The first is the size of the rank two opinion that cannot be too small compared to the size of rank one opinion. The second condition requires that the size of rank one opinion should not be too large a fraction of the entire population of the condition. 
When one of these two inequalities hold, the rank switching happens. More formally, let $s_{i}^p$ denote the number of participants in $O_i^p$ opinion, and $N_{CS}$ denote the size of the relevant group of condition $CS$. The first inequality limits the size of $s_1^1$ in terms of absolute population size of the condition, $s_1^1<2N_{CS}/3$, while the second limits is expressed in terms of the size of the second rank population as $s_2^2\geq s_1^1/4$. Both criteria are sufficient and combined they are
necessary, and each eliminates two out of three inkblots preserving the ranking. We believe that those criteria use universal features but require for each system adjusting experimentally coefficients in the two inequalities. 

\subsection*{Effects of confidence on the impact of ambiguous influence}
In addition to providing an opinion per inkblot in our study, participants were asked to rate their confidence in their opinions along a four-point scale. Using confidence ratings collected in the first phase, we analyze the frequency of participants conforming with the promoted opinion in the CM and CS conditions as a function of two arguments. The first argument is the participant's confidence level from phase one. The second is a binary value 1 when the participant's opinion in phase one was the promoted one, and 0 otherwise.

\begin{figure}
    \centering
    \begin{subfigure}[h]{0.85\textwidth}
        \centering
        \caption{}
        \label{fig:promoted_vs_confidence}
        \includegraphics[width=0.75\textwidth]{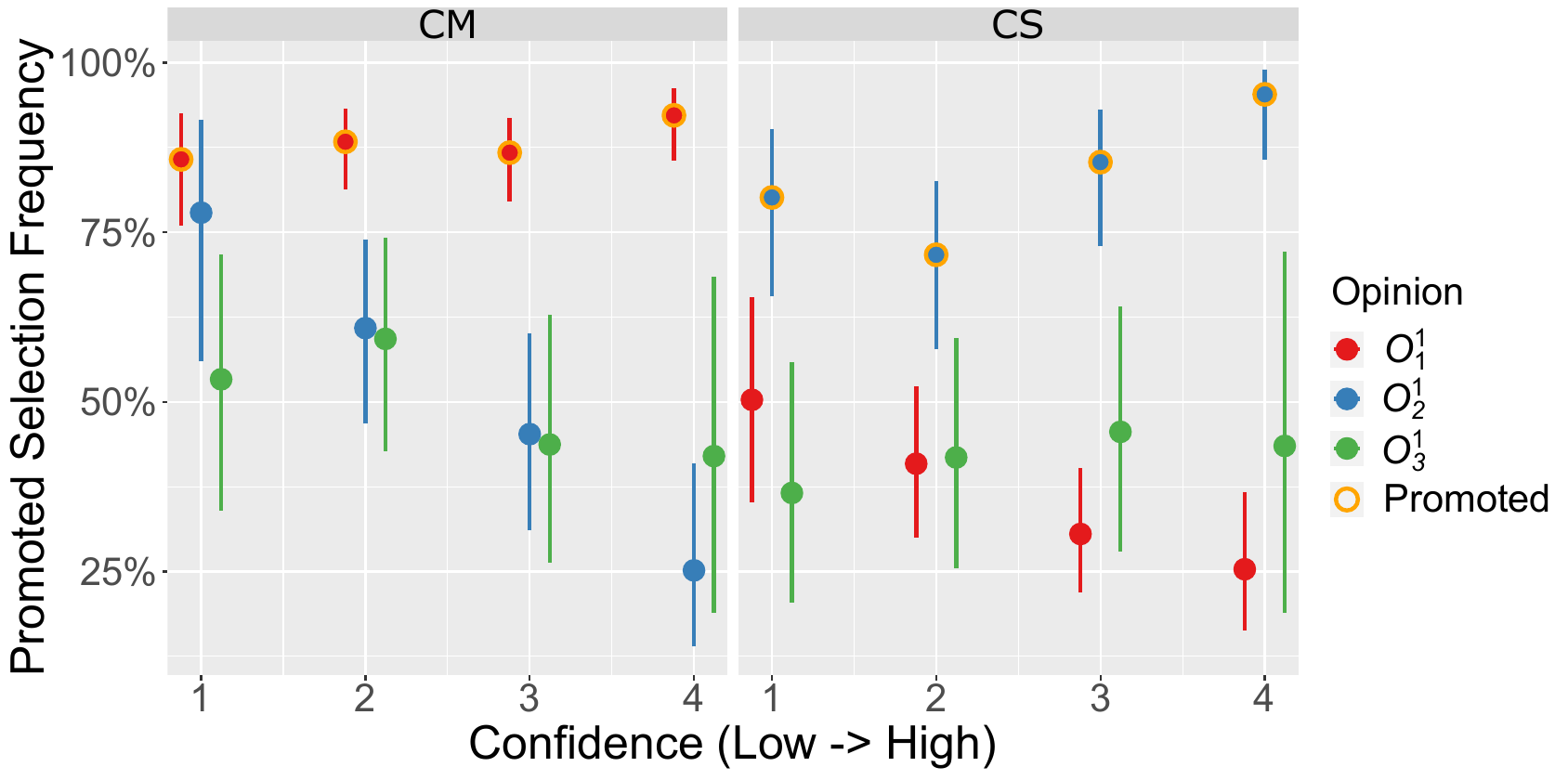}
    \end{subfigure}\\
    \begin{subtable}[h]{0.85\textwidth}
        \centering
        \caption{}
        \label{tab:chi_squared}
        \begin{adjustbox}{width=\columnwidth,center}
        \begin{tabular}{l|l|l|l|l|l|l|l|l|l|l}
    	\hline
            \textbf{Inkblot} & 2 & 1 & 10 & 4 & 7 & 5 & 8 & 6 & 3 & 9 \\ \hline
            \textbf{Entropy} & 1.03 & 1.235 & 1.312 & 1.413 & 1.451 & 1.484 & 1.561 & 1.581 & 1.719 & 1.799 \\
            \textbf{p-value} & 1E-24 & 0.0005 & 0.001 & 3E-07 & 0.0005 & 0.327 & 0.831 & 4E-05 & 0.831 & 0.774 \\
        	\textbf{Reject $H_0$?} & Yes & Yes & Yes & Yes & Yes & No & No & Yes & No & No \\ \hline
    	\end{tabular}
    	\end{adjustbox}
    \end{subtable}\\
    \caption{\textbf{Participant confidence and effects on conforming to the promoted opinion.} (\subref{fig:promoted_vs_confidence}) shows how confidence impacts the selection of the promoted opinion for each phase one opinion in the CM and CS conditions. (\subref{tab:chi_squared}) shows a chi-squared analysis of the confidence ratings for each inkblot. Each inkblot has the same data sample size, $N_{p2}=242$. Each column represents the results of an inkblot's chi-squared test run on the inkblot's answers and a tally of the participant's opinions on that inkblot's representation, categorized by the four confidence levels.}
    \label{fig:confidence_analysis}
\end{figure}

Figure~\ref{fig:promoted_vs_confidence} shows the results of our regression model, which estimates the frequency of conforming with the promoted opinion in phase two as a function of the interaction between the participant's condition in phase two (either CM or CS), the opinion that participant held in phase one, and the confidence the participant had in their phase one opinion. This means that we compute a frequency estimate for each combination of the above factors. We chose to limit the analysis to the CM and CS conditions because, unlike the C and CR conditions in which each participant holds in phase one opinion promoted in phase two, their participant's phase one opinion may be promoted or not in phase two.

This figure reveals that there is a strong linear decrease in the selection of the promoted opinion for N participants as the confidence level increases. This negative trend holds for N participants holding the popular opinions ($O^1_1$ for CS and $O^1_2$ for CM). The sparsely populated $O^1_3$ has little discernible trend. We find that $O_3^1$'s small population size (roughly half the size of $O_2^1$) contributes the most to the absence of trend and wide credible intervals shown by the model. Additionally, for the CS condition, we see a strong linear increase in promoted opinion selection for P participants as confidence level increases. We discuss more details on this analysis, and the underlying model used to create this figure, in SM Section 1.4.1.

In conclusion, as expected, the strength of the participants' initial confidence in their selection of opinions correlates with their reluctance to change opinions. A participant has significantly more confidence in their selection of opinion if it is promoted by the ambiguous comments shown to participants in CM and CS conditions. The confidence level of participants defines their vulnerability to ambiguous influence, with high confidence levels translating into a low likelihood to succumb to influence.

\subsection*{Relationship between content complexity and confidence}
The task of quantifying the complexity, or ambiguity, of an inkblot and the associated opinion is difficult, as interpreting such content is a visual task. To keep it simple, we use base 2 Shannon entropy \textit{(32)} to measure the fractions of participants that hold each opinion for each inkblot, see Methods for details. This measures confusion of participants as a function of the distribution of opinions that result from interpreting an inkblot, with a value of $0$ when all participants selected one opinion, to a maximum value of $2$ when all opinions are chosen by the same number of participants. See Supplemental Figure S1 and Fig.~\ref{tab:chi_squared} for a list of the inkblots and their chosen opinions entropy values.

We performed a chi-squared analysis on the observed frequency of the first phase confidence ratings for the opinions of each inkblot. In this test, if the chi-squared null hypothesis, $H_0$ is rejected for a particular inkblot with $p < 0.05$, then there is a clear relationship between an opinion and the level of confidence participants assigned to it. This would indicate that participants felt inherently confident about choosing such an opinion in the second phase, implying its popularity. A failure to reject $H_0$ indicates a lack of relationship between an opinion and the confidence level participants assigned to it, implying that the inkblot and its opinions are not easy to interpret. We run this test on every inkblot and tabulate the results in Fig.~\ref{tab:chi_squared}.
 
The results are sorted by the opinions' entropy values, showing that a cut-off for the rejection of $H_0$ for entropy values is larger than $1.451$, except for inkblot 6. This means that at a certain level of entropy, participants lose confidence in their selections of opinions. As mentioned previously, lower confidence makes these participants more susceptible to ambiguous influence. This conclusion combined with the observations shown in this section, implies that increasing the content complexity, i.e., ambiguity, makes the recipients more susceptible to ambiguous influence.

\section*{Discussion}
 
The main expected result of this paper was pre-registered as follows: "...given information about the opinions of others they (participants) will be likely to conform to opinions consistent with the presented information beyond the natural rate of opinion transitioning frequencies without such external information." Our results validate this hypothesis and using them we discover more correlations. The opinions with unknown origins exert a meaningful influence on participants that may increase the rank of the promoted opinion which is not the most-popular opinion. When the second most popular opinion is promoted, in 70\% of such cases the outcome of opinion promotion significantly distorts the opinion's popularity by switching the ranks of two popular opinions. 

We also found that the CR condition combining both internal influences only slightly enhances preservation of the initial opinion.
The reason is that each participant in CR sees their own single comment about the opinion currently held by this participant. Thus, most participants in CR will see a comment that has lower quality than each of the five comments seen by participants in CM and CS conditions. The two external influences rely on partial information to encourage participants to select the promoted opinion in the second phase regardless of whether this requires changing or keeping the opinion. The fraction of participants doing so measures the magnitude of this influence. 

We also found that large differences in participants' frequencies of changing initial opinions are negatively correlated with the strength of peoples' confidence in those opinions. However, confidence itself depends on the complexity of the content of an inkblot and its opinions. Using entropy, we found that the more complex the content is, the less confident the participants are about their chosen opinion, making them more susceptible to external influence.  This implies that bad actors seeking to spread influential misinformation may intentionally obfuscate information to reduce users' confidence in their own opinions, causing more users to follow the bad actor's influence.

Historically, the influence through human interactions was controlled by body language, which could trigger warnings against even familiar influencers,  and certainly unfamiliar ones with unknown positions in society. From that perspective, our results are significant because they demonstrate that online influence can be triggered by comments authored by  the introduction of unknown participants, opening the gates for unrestricted online influence.

With the growth of misinformation both proliferated and generated by bots \cite{shao2018spread,wang2018era,himelein2021bots,zellers2019defending,de2023chatgpt}, concerns must be raised on how even the passive viewing of online information with unknown provenance can influence people into changing  their opinions, especially when that information is tied to polarizing subjects \cite{flamino2023political}. Such influence is easier to create with the introduction of increasingly accessible generative Artificial Intelligence (AI) models designed for text generation and other media. These AI models increase the danger that deceptive or harmful information can be easily created and spread to influence masses. 

In future research, we will expand upon the foundation of analysis presented here and explore how the frequency of users changing opinions varies with different kinds of content and mediums. Another important direction of this research is to explore the observation that increasing inkblot complexity led to statistically weaker participant confidence. Hence, we will study if such a threshold exists in other mediums. Finally, we are also interested in exploring if simulated elections could be manipulated by misinformation by promoting the candidate who is second in the polls.

\section*{Acknowledgments}
BKS and BC were partially supported by NSF Grant No. SBE-2214216 and DARPA-INCAS Grant under Agreement No. HR001121C0165. 

\section*{Author Contributions}
Conceptualization: BC, BKS, JH; Data curation: BC; Formal Analysis: BC, BKS, JB; Funding acquisition: BKS; Methodology: BC, BKS, JB, JF, JH; Project administration: BKS; Resources: BKS;
Software: BC; Supervision: BKS; Validation: BC, BKS, JB, JF;
Writing – original draft: BC, BKS.
All authors participated in investigation and writing-editing-reviewing the paper.

\section*{Competing Interests}
The authors declare no competing interests.

\section*{Data and Materials Availability}
After the paper is accepted we provide access to code and data.

\section*{Supplementary Materials}
Materials and Methods\\
Supplementary Text\\
Figs. S1 to S13\\
Tables S1 to S9\\
References \textit{(29-32)}

\end{document}


\pagenumbering{arabic} 

\maketitle

\baselineskip24pt
\textbf{These Supplementary Materials include:}

\indent \indent
Methods

\indent \indent
Supplementary Figures S1 to S13

\indent \indent
Supplementary Tables S1 to S9
\newpage

\section{Methods}
\label{sm_part:materials_and_methods}

\subsection{Study Deployment and Participants' Demographic}
 
The two phases of our survey were conducted over the course of eight weeks during the months of July, August and September of 2022. Phase 1 of our experiment was delivered to Prolific users on  07/18/2022 at 12pm and lasted for two days until 07/20/2022 at 1pm. After collecting the initial 300 responses, we spent the following two weeks processing the data from these initial responses, assigning participants to experimental conditions, and setting up the different surveys. After this preparation phase, we released the second phase of our study to users on 08/01/2022 at 12pm. We released the second phase into two groups to check that there was no issue with the surveys or resulting data being collected. Phase 2 lasted until 09/07/2022, with 92\% of the data having been collected in the first two weeks from the time  the second phase being released. We decided to allow the study to stay up for an additional few weeks to get as much data as possible for analysis.

In terms of demographics of the 300 initial participants, all participants' country of residence was the United States, and all identified as speaking fluent English. The average age was 45.7 years,  with a standard deviation of 15.8 years. 146 participants identified as Male and the other 154 identified as Female. For ethnicity, 223 participants identified as White, 38 as Black, 19 as Asian, 10 as Mixed, and 10 as Other.

\subsection{Survey and Experimental Design}
\label{sm_section:survey_and_experiment_design}

In this section we will give additional information on the study and survey design of this research, including examples of the surveys sent to participants in both phases and of inkblots with different opinions and levels of entropy. We also provide details of assignment of participants to conditions in the second phase.

We start by showing some examples of inkblots and the options participants could select from, shown in Figure \ref{sm_fig:inkblot_difficulties}. We also show the entropy of these inkblot's opinion selection distribution, which gives an idea of how complex the inkblot was for participants. In section \ref{sm_section:exploratory_analysis} we discuss the topic of inkblot entropy in more detail.

\subsubsection{Survey Example inkblots}
\label{sm_section:survey_examples}
\begin{figure}[!ht]
    \centering

    \begin{subfigure}[t]{0.33\textwidth}
        \centering
        \includegraphics[scale=.45]{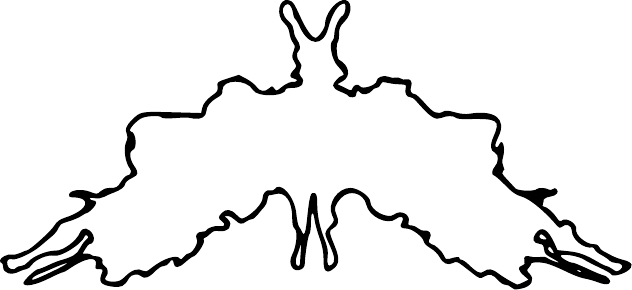}
    \caption{\textbf{Low Entropy}}
    \label{sm_fig:low_difficulty_inkblot}
    \end{subfigure}%
    \centering
    \begin{subfigure}[t]{0.33\textwidth}
        \centering
        \includegraphics[scale=.45]{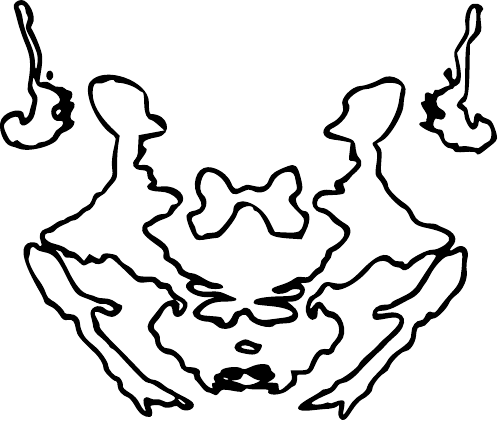}
    \caption{\textbf{Medium Entropy}}
    \label{sm_fig:medium_difficulty_inkblot}
    \end{subfigure}
    \centering
    \begin{subfigure}[t]{0.33\textwidth}
        \centering
        \includegraphics[scale=.45]{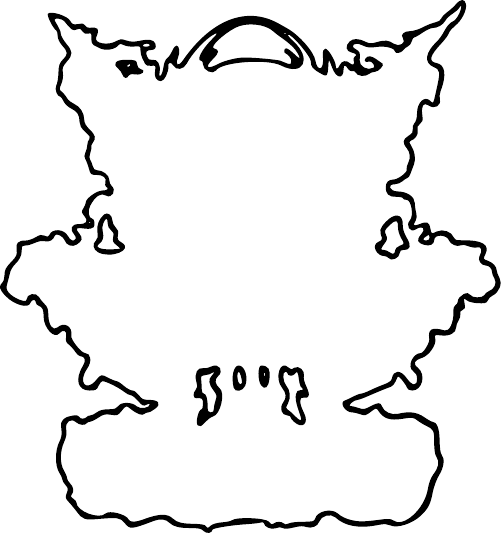}
    \caption{\textbf{High Entropy}}
    \label{sm_fig:high_difficulty_inkblot}
    \end{subfigure}
    \begin{adjustbox}{width=1\columnwidth}
        \begin{tabular}{|lllllll|}
        \hline
        \multicolumn{1}{|c}{\multirow{2}{*}{\textbf{Inkblot}}} & \multicolumn{1}{c}{\multirow{2}{*}{\textbf{Entropy}}} & \multicolumn{1}{c}{\multirow{2}{*}{\textbf{Avg. Confidence}}} & \multicolumn{4}{c|}{\textbf{Opinion (pop. Frac,   avg. conf)}}                                               \\
        \multicolumn{1}{|c}{}                         & \multicolumn{1}{c}{}                         & \multicolumn{1}{c}{} & \textbf{1 (Most Popular)}& \textbf{2}& \textbf{3}& \textbf{4 (Least Popular)}\\ \hline
        (a) Low Entropy & 1.03 & 3.42 & Bat (.537, 3.36) & Butterfly (.458, 3.52) & Scissors (.004, 1.0) & Crocodile Heads (0, N/A) \\
        (b) Medium Entropy                                         & 1.45                                         & 2.5                                                  & Two Figures (.55, 2.29) & Two Women (.33, 2.90)  & Ribbon (.058, 2.0)   & Two Men (.049, 2.75)     \\
        (c) High Entropy                                         & 1.79                                         & 2.33                                                 & Explosion (.40, 2.30)   & Map (.34, 2.32)        & Flower (.15, 2.36)   & Man (.09, 2.47) \\ \hline
        \end{tabular}
     \end{adjustbox}
    
\caption{\textbf{Inkblot examples at different entropies}. This figure shows three of the inkblots presented in our surveys. For each inkblot we give the entropy for the first phase opinion counts, the average confidence of the inkblot, and the list of opinions the participants could choose from. The average confidence is represented numerically by encoding the confidence categories into values 1 - 4 with 1 being the least and 4 being the most confident. For each opinion, we provide the fraction of participants who chose the opinion as well as their average confidence. The opinions inherit their rank from the place of their popularity on the list of all opinions' popularity's ordered from the highest to the lowest.}
    \label{sm_fig:inkblot_difficulties}
\end{figure}

Examples of three inkblots from the study are shown in Figure \ref{sm_fig:inkblot_difficulties}. To see the differences in what participants saw in each of the different conditions of the experiment, see Figures S11-S14.

\subsubsection{Influencing Comment Selection and Presentation}
The main factor affecting the change of influence strength among the different experimental conditions is 
the various number and quality of comments shown to the participants under these conditions. While many comments left by participants were descriptive, their quality varied widely. To reduce potential confusion from low quality comments, for CM and CS, but not for CR, we hand-selected comments according to the quality of their level of details and uniqueness of description of the inkblot. For each inkblot and each of its popular opinions, the authors selected the six high quality comments. If a participant in CM and CS wrote a comment included in the top five selected comments, the participant was shown the remaining five comments. Otherwise, the five comments judged the most specific to the corresponding inkblot were shown.

\subsubsection{Condition Prompts and Opinion Transitions}

The participants in CR condition are shown the following prompt along with their own comment from the first phase: \textit{"Please read the following comment 
about the above image before choosing the best description below"}. The participants in CM and CS are shown comments supporting opinion X which stands for either the most popular (CM) or second-most-popular (CS) opinion from the first phase, along with the following text: 
\textit{"Please read the following comments about the choice X, one of the popular choices in stage 1"}.

\begin{description}
\item{{Confirmation bias (C):}} promotes three opinions $O_{r=1,2,3}^r$ and the P subgroup includes participants with three transitions $O_{r=1,2,3}^1\rightarrow O_r^2$, while the N subgroup consists of participants with six transitions 
$O_{r=1,2,3}^1\rightarrow O_{r\neq s=1,2,3}^2$. 

\item{Reinforced Confirmation Bias (CR):} reinforces the influence about the same three opinions $O_{1,2,3}^1$ as in C, by showing each participant their comment from the first phase. Therefore, the corresponding transitions for the P and N subgroups as for the C condition. Meaningful comments will increase participant reluctance to choose a different opinion above the reluctance caused by the confirmation bias alone.

\item{The most-popular opinion influence (CM), and the second-most-popular opinion influence (CS):} Let $P$ be 1 for CM and 2 for CS. There are three promoted opinions for each participant, one defined by comments and a note $O_P^2$ with transition $O_P^1\rightarrow O_P^2$ also supported by confirmation bias, and two more defined by confirmation bias with transitions $O_{r=3-P,3}^1\rightarrow O_r^2$, so the total of five transitions defining P subgroup, and remaining four, defining the N subgroup.
\end{description}

\subsubsection{Assignment of Participants to the Second Phase Conditions}
\label{sm_section:p2_condition_assignment}
After the conclusion of the first phase, our next task was to assign participants to the four conditions of the second phase. We decided to equally assign the participants to each condition ahead of sending out the second phase surveys, so this meant randomly splitting our 300 participants into four equal 75 participant groups. The participants were assigned to conditions in such a way that it would minimize the difference between the first phase opinion selection distribution of each condition versus the overall distribution. However, assigning participants ahead of time opened up the possibility for an unbalanced group for our different conditions if not all participants responded to the second phase surveys. 

An analysis of the likelihood of different condition group distributions resulted in our ideal 60, 60, 61, 61 distribution having a $\sim 1.3\%$ chance of occurring, and the chance of us having a reasonable balance (difference of at most 15\% between highest and lowest bin) to be $\sim 80\%$ and the chance of being poorly balanced (\textgreater{} 25\% difference between highest and lowest bin) to be less than 2\%. We determined that the risk of a poorly balanced the second phase condition assignment was low enough so that we could assign participants to conditions in phase two, instead of a more complicated approach like assigning as the participants return to the second phase. Additionally, this decision was informed by how we delivered the survey, as a link to a google form document provided to users through Prolific.

\subsection{Modeling and Analysis}
In this section we discuss the various alternative models we used to examine our data. We first discuss the second phase opinion selection model which is used in the main text, and then we discuss alternative models that examine the problem from a different perspective. The results of all models shown all agree and support the conclusion that ambiguous comments have a strong ability to convince participants to conform with their promoted opinion.

\subsubsection{Phase Two Opinion Selection Model}
For our regression model we used a Bayesian multilevel multinomial regression model. Its goal is to predict the probability of each opinion being selected in the second phase, based on the data from the first phase. We use these predicted opinion selection probabilities to determine if the external influence cause a significant increase in selection probability for their corresponding promoted opinion, versus control. To model this we use opinion transition data, where each data-point contains some participant's opinion for an inkblot in the first phase,  its corresponding confidence, comment, and the opinion selected for the same inkblot in the second phase. This gives us a data-set of 10 times the number of participants in our study.

As suggested previously, the dependent variable for this model is the fraction of participants who selected opinions $O^2_i$, where $i \in {1,2,3} $. The independent variables used in the best fitting model are: (i) Condition dummy variable indicating the experimental condition under which the participant was a member in the second phase. (ii) Phase 1 opinion dummy variable indicating the opinion selected in the first phase. This variable is not ordered, for reasons we discuss in the model selection portion of the SM. The regression also models participants and inkblots as random effects to account for the repeated measures taken by each participant and potential group-wise similarities of the inkblots.

\subsubsection{Model Comparison and Selection}
\label{sm_section:p2_opinion_selection_model}

To select the best model for predicting users’ choices of opinions based on experimental conditions, we compared models using Leave One Out (LOO) cross-validation and performed the selection steps [29]. To do this, we iterated through a set of potential variations of the model, increasing the number of explanatory variables and how the interaction of confidence and $O^1_{1,2,3}$ and $O^2_{1,2,3}$ were modeled until we
reached the strongest candidate model. In all, we tested eight different models displayed in table S1 in which we define increasingly complex regression models aimed at explaining how opinions are chosen
in the second phase. The null model represents this transitioning as influenced only by the participants themselves. The following models are additive; they increase the parameters of the model to include more experiment information.

\begin{table}[hp!]
    \centering
    \resizebox{.7\textwidth}{!}{
    \begin{tabular}{l|cc|c}
        & \multicolumn{2}{c|}{LOO ELPD} &\\
        Model & diff & s.e. diff & Model Weight
        \\ \hline
        Model.null                  & -513.7    & 31.3  & 0.0  \\
        4 Condition                 & -484.8    & 30.9  & 0.004 \\
        Inkblot + rank              & -1.8      & 6.4   & 0.413 \\
        Inkblot + rank (ordinal)    & -210.7    & 21.2  & 0.020\\
        Simple Cond. x rank         & 0.0       & 2.7   & 0.033\\
        P1 Confidence               & -2.4      & 6.2   & 0.001 \\
        P2 Confidence               & -2.5      & 5.9   & 0.000 \\
        Condition x Rank            & 0.0       & 0.0   & 0.496 \\
    \end{tabular}}
    \caption{\small{\textbf{Bayesian candidate model comparison.}
    This table compares the set of the second phase opinion selection models. The null model represents opinion purely as a random participant factor. All models following the null model are built on the model before it, adding the terms described in the model column. The second two columns show the results of a pairwise comparison of the Bayesian LOO estimate of the expected log point-wise predictive densities [30] where 0.0 indicates the strongest model and all other models are compared with it. The final column shows the weight of each model, where the weight is determine by combining all the models and finding the optimal weights for this combination. Models with 0 or very low weights are not very significant. These methods are implemented in the 'loo' R package [31]. The Condition x rank models represent replacing the condition and baseline opinion rank variables with an interaction term between them. This allows us to model an intercept per condition / rank combination, giving us flex-ability to model rank specific effects per condition. The Simple Cond. x rank model is the Inkblot+rank model with this interaction replacement. The strongest two models shown are the Inkblot + rank model and the Condition x Rank model. The simpler model does nearly as well as the more complex model, and thus to avoid overfitting we use the simpler model in the main text.
    }}
    \label{sm_tab:bayes_anova_model_compare}
\end{table}

The most important modeling decision was to properly represent the 
$O^{1,2}_{1,2,3}$ terms in our regression. As we have discussed, an opinion’s transition rank correlates with the opinion’s ability to attract and retain participants. Since each influencing condition promotes specific opinions, we need to allow the influence to override the effects of opinion rank to which this participant transitions. To do this, we model the interplay between condition and phase of an opinion rank, allowing for variation across them. 

It is also important to determine how to numerically represent the opinion rank in our regressions. It is common to model ranks using ordinal variables, as there is an inherent relationship between any two ranks. However, we could have instead chosen the fraction of participants who selected this opinion. Such a model would have allowed us to vary the distance between opinions based on the distance between these fractions, instead of uniformly spreading opinions across ranks. Both are good approaches, but they fail to account for the influence portion of the experiment. When influence promotes specific opinions, it reverses the relationship that opinion rank had with transitioning frequency. The CS condition, for example, has inverted selection frequencies for 
$O^2_1$ and $O^2_2$ than one would expect from an ordinal variable, since $O^2_2$ is the promoted opinion for the CS condition and it is subsequently selected more frequently than $O^2_1$. If this were modeled by ordinal or fractional sizes of popularity, not opinion ranks, we would be constraining our parameter space by enforcing some consistent relation between all opinion ranks that do not exist when influence is applied. This is reflected in model comparisons. The 
most complex model with 21 parameters had a significantly better (penalized) fit ($p<0.001$) than simpler models.

\subsubsection{Transition Frequency Regression Model}
In addition to modelling what informs participants to select particular opinions in the second phase, we also modeled the factors that trigger the participant to transition their opinion away from the one they held in the first phase. This transitioning frequency gives us an idea of how ambiguous comments may strengthen or weaken their held opinions from the first phase. 

Table S2 shows the results of the regression model. It includes both the condition and the first phase
opinion as nominal variables and the interactions between both terms. Both conditions and opinions have reference levels, the control condition and opinion $O^1_1$. This makes the remaining conditions or rank parameter estimates relative to the control opinion. So each listed condition’s parameter estimate shows how
each condition transitioning frequency differs from control $O^1_1$. The opinion rank parameter estimates show how opinions $O^1_1$, $O^1_2$ and $O^1_3$ differ in transitioning rate from the control opinion.
\begin{table}[hp]
	\Large
	\centering
    \resizebox{.9\textwidth}{!}{
	\begin{tabular}{|l r r r|}
	\hline
    	Predictors & Odds Ratios & CI & p-value \\
    	\hline
    	Condition (C) $O_1^1$ (Intercept) & 1.37 & 0.70 - 2.69& 0.358\\
    	Inkblot & 0.97 & 0.91 - 1.04 & 0.399\\
        Condition (CR)& 0.61 & 0.39 - 0.95 & \textbf{0.030}\\
        Condition (CM)& 0.52 & 0.33 - 0.82& \textbf{0.005}\\
        Condition (CS)& 2.61& 1.74 - 3.92& $\mathbf{<}$\textbf{0.001}\\
    	$O_2^1$ & 2.49& 1.60 - 3.89& $\mathbf{<}$\textbf{0.001}\\
        $O_3^1$ & 4.32& 2.61 - 7.51& $\mathbf{<}$\textbf{0.001}\\
        Phase 1 Confidence & 0.74 & 0.66 - 0.84& $\mathbf{<}$\textbf{0.001}\\
    	Phase 2 Confidence & 0.81& 0.72 - 0.92& \textbf{0.001}\\
        Condition (CR) - $O_2^1$& 1.02& 0.54 - 1.94& 0.940\\
        Condition (CM) - $O_2^1$& 3.13& 1.66 - 5.93& $\mathbf{<}$\textbf{0.001} \\
        Condition (CS) - $O_2^1$ & 0.12 & 0.06 - 0.22 & $\mathbf{<}$\textbf{0.001}\\
        
    	Condition (CR) - $O_3^1$ & 0.96& 0.46 - 2.04& 0.926\\
        Condition (CM) - $O_3^1$ & 2.98& 1.42 - 6.26& \textbf{0.017} \\
        Condition (CS) - $O_3^1$ & 0.64& 0.31 - 1.32& 0.230\\ \hline
     \end{tabular}
     \begin{tabular}{|l|l|}
     \hline
        \multicolumn{2}{|c|}{Random Effects} \\
    	\hline
        $\sigma^2$ & 3.29 \\
        $\tau_{00\ user}$ & 0.38\\
        $\tau_{00\ inkblot}$ & 0.08 \\
        ICC & 0.12 \\
        $N_{user}$ & 242 \\
        $N_{inkblot}$ & 10 \\
        \hline
        \hline
    	Numerical Observations & 2420 \\
    	Marginal $R^2$ & 0.201\\
    	Conditional $R^2$ & 0.299\\
	\hline
	\end{tabular}}
  \caption{\small{\textbf{Multilevel logistic regression of opinion transitioning frequencies}. 
    	This table defines the parameters and coefficients of our mixed effects model. The columns, in order, define the parameter name, the effect on transitioning frequency each parameter has (as an odds ratio), the confidence interval of the odds ratio estimate, and the p-value of the parameter in the model. The significant parameters have their p-values shown in bold. The dependent variable for the regression model is frequency of opinion transitioning. The independent variables are the conditions, phase one opinions ($O_i^1$)}, two confidence levels, and the inkblot id. Each participant selects one opinion for each of 10 inkblots in the first phase and keeps or changes it in the second phase, creating $242\times 10$ trials for frequency of opinion transitions. This model accounts for designed effects of conditions and groups as well as for the random effects caused by each participant having 10 trials to change or keep opinion for each inkblot. The model coefficients are presented as odds ratios, where the control group is the reference group for the other conditions, two of which are split into subgroups resulting from influence information provided to participants in these conditions. The opinion rank similarly has a reference level, which is the most popular opinion in the first phase. The coefficients show the strength of impact that each condition and subgroup had on the opinion transitioning frequency.  The p-values of all important terms are significant ($p < .05$).}
    \label{sm_tab:experiment_vs_movement_regression}
\end{table}

\begin{figure}[hp]
\centering
\includegraphics[width=0.95\textwidth]{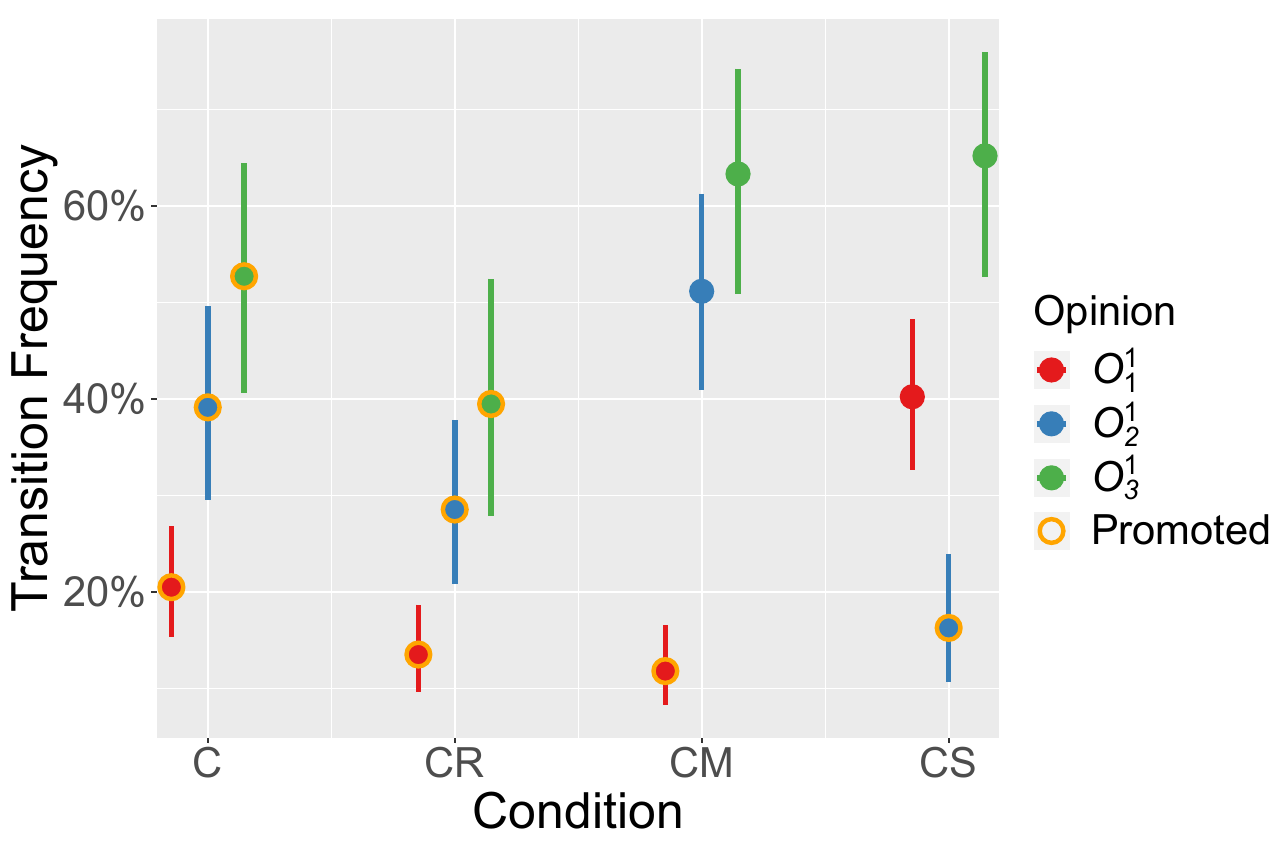}
\caption{\textbf{Predicted opinion transitioning frequency by condition}.
    This figure shows the results focusing on the relationship between condition, opinion, and predicted transitioning frequency. In this figure we can see that each influencing condition has some significant effect on reducing the amount of transitioning that occurs for participants holding the promoted opinion of the condition in the first phase.}
\label{sm_fig:predicted_opinion_transition_frequency}
\end{figure}

\begin{figure}[hp]
    \begin{center}
    \begin{subfigure}[t]{0.45\textwidth}
    \centering
        \includegraphics[width=0.9\linewidth]{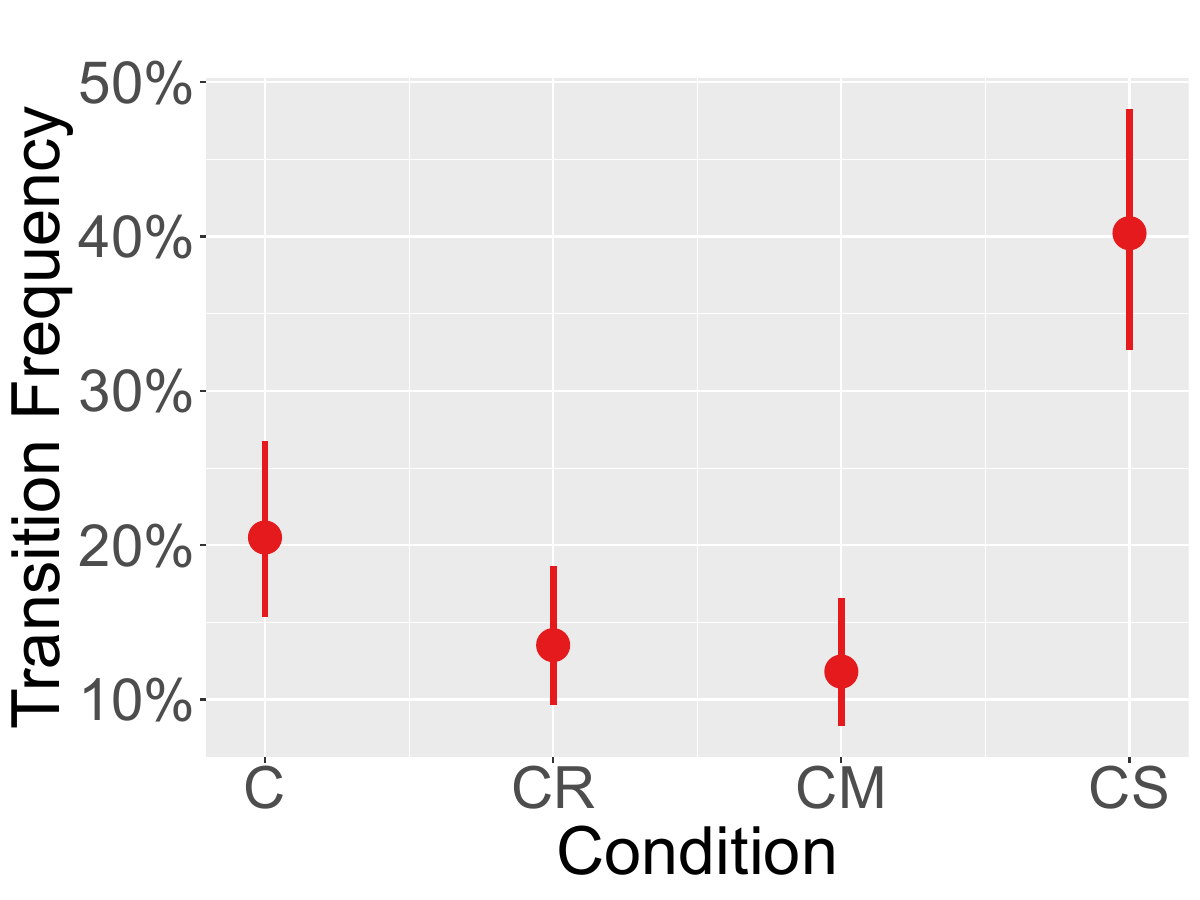}
    \end{subfigure}%
    \centering
    \begin{subfigure}[t]{0.45\textwidth}
    \centering
        \includegraphics[width=0.9\linewidth]{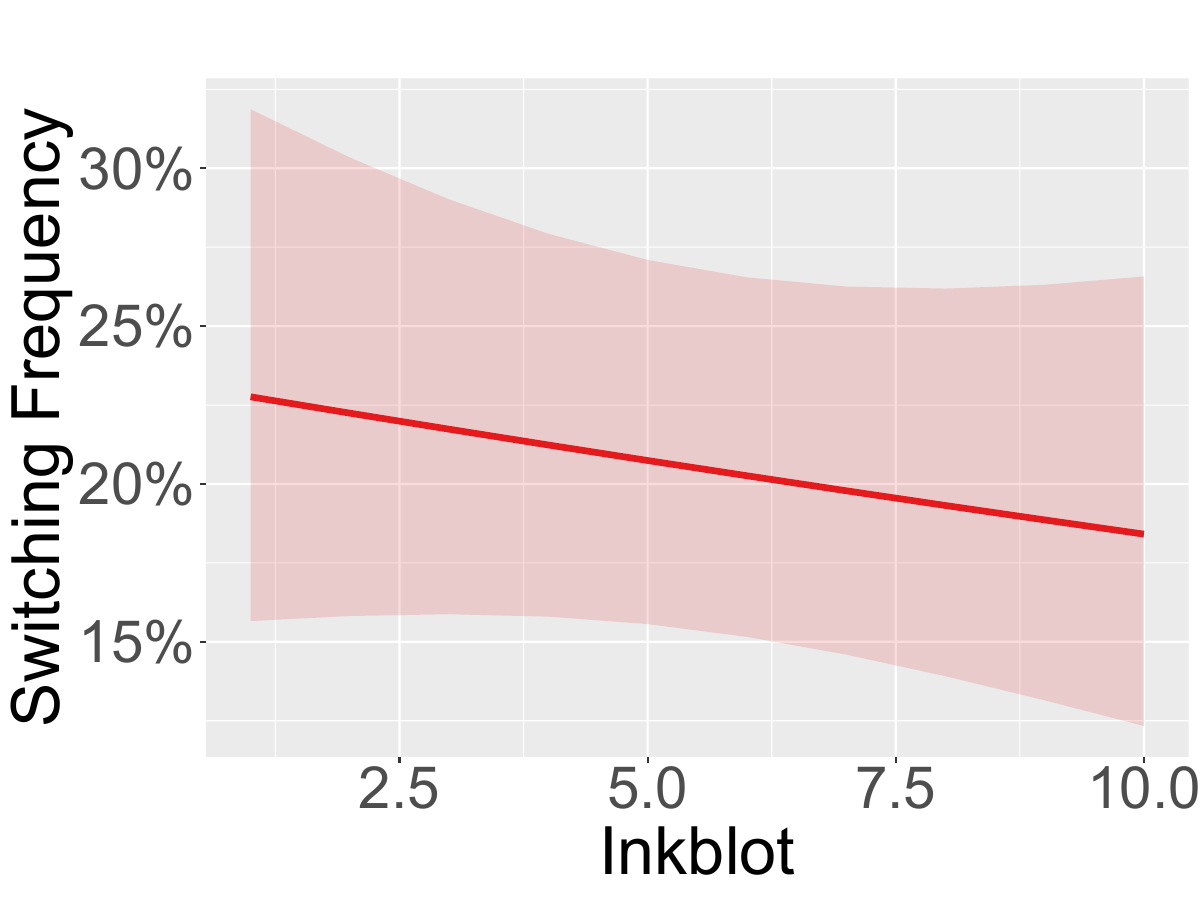}
    \end{subfigure}
    
    \begin{subfigure}[t]{0.45\textwidth}
    \centering
        \includegraphics[width=0.9\linewidth]{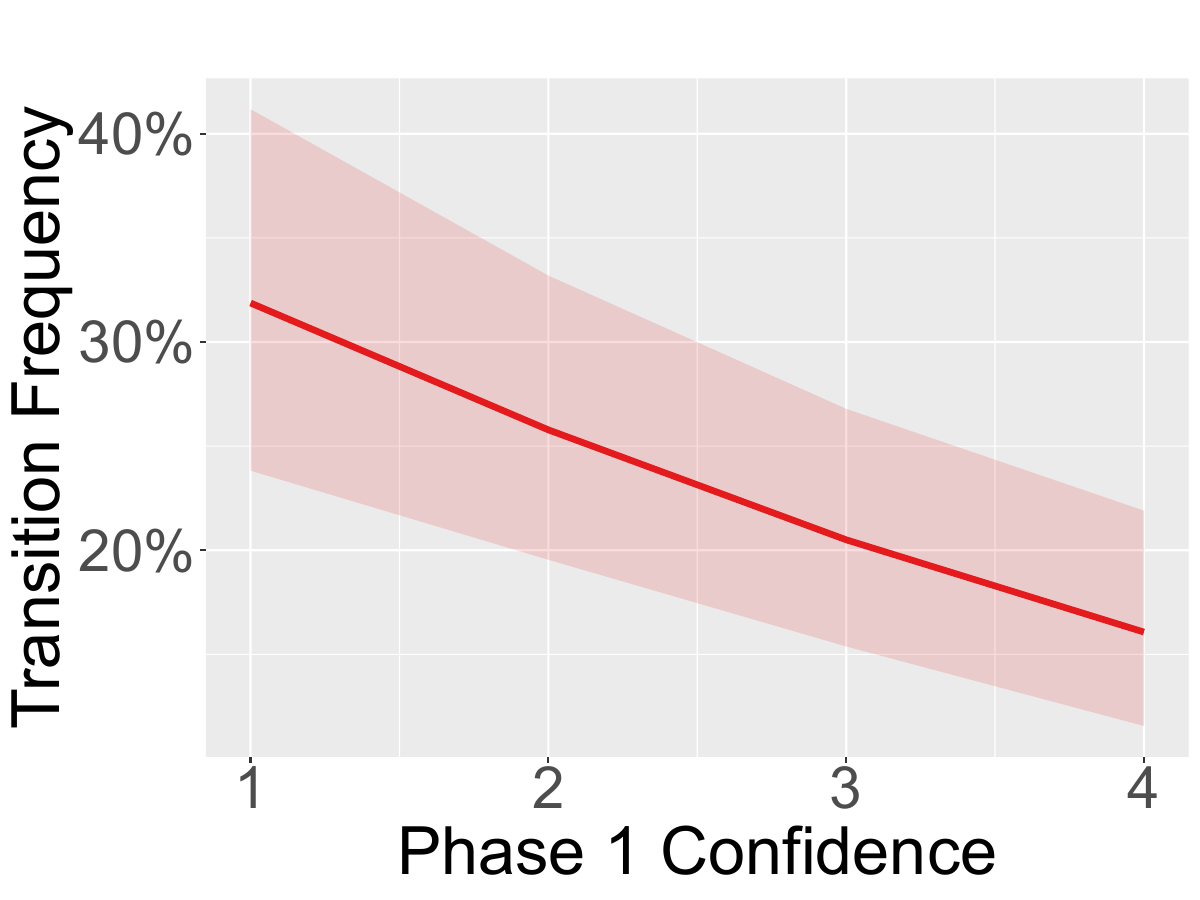}
    \end{subfigure}%
    \centering
    \begin{subfigure}[t]{0.45\textwidth}
    \centering
        \includegraphics[width=0.9\linewidth]{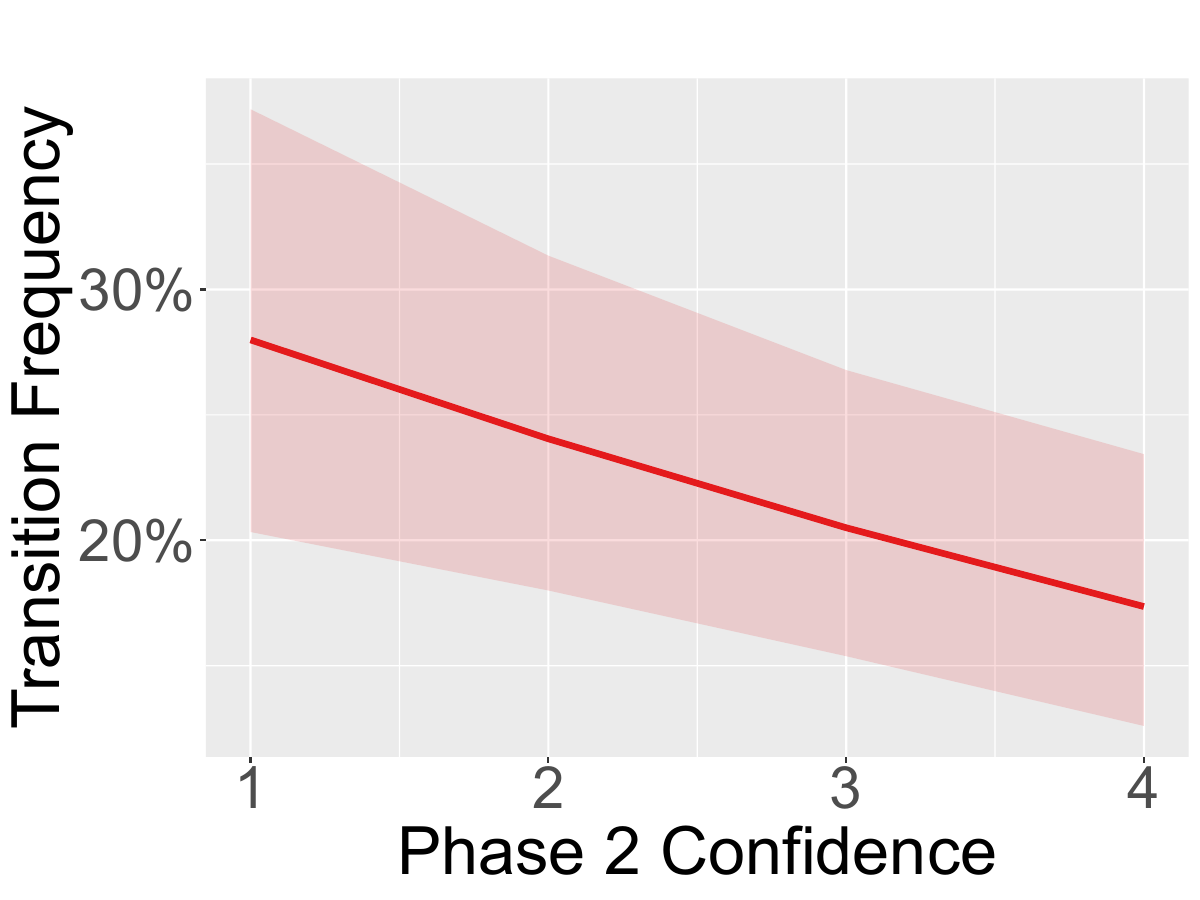}
    \end{subfigure}

        \begin{subfigure}[t]{0.45\textwidth}
    \centering
        \includegraphics[width=0.9\linewidth]{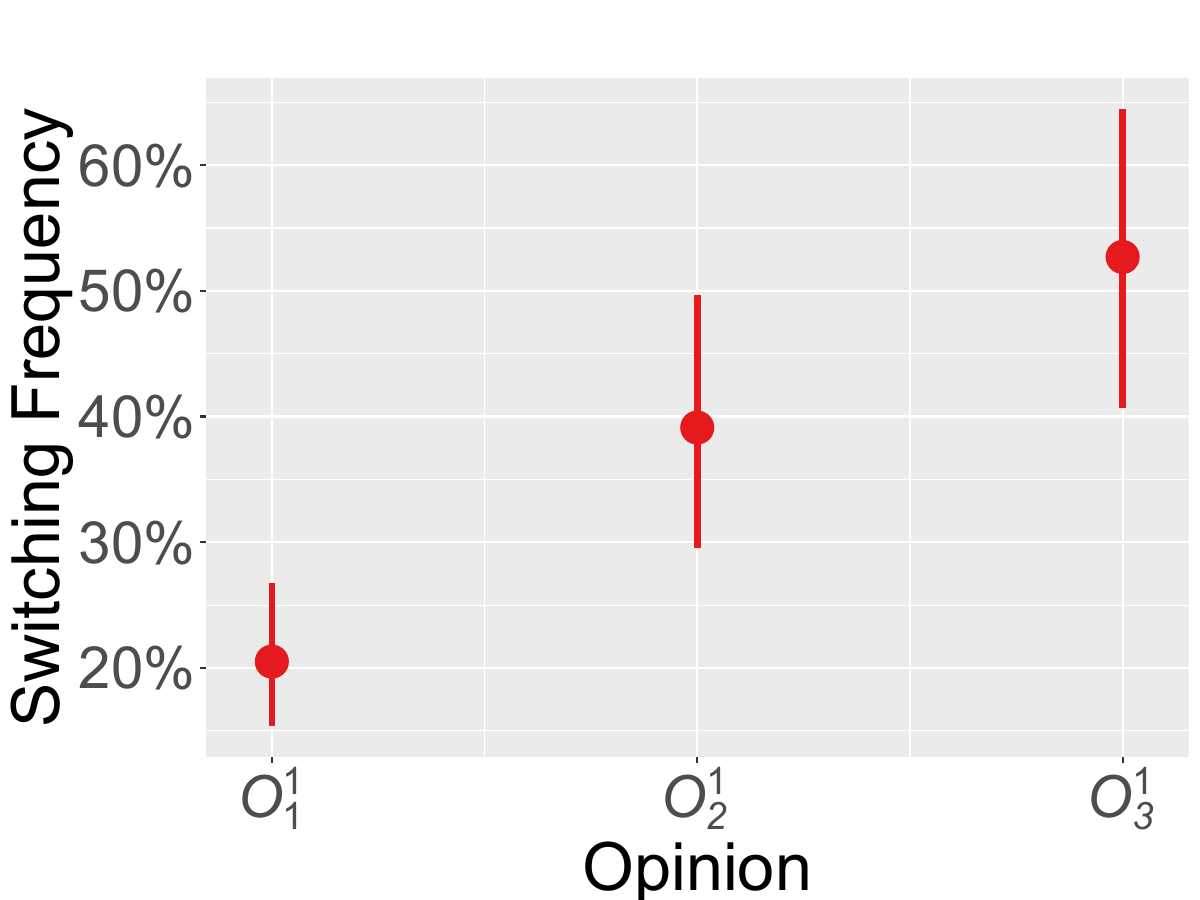}
    \end{subfigure}
    
    \caption{Categorical Regression Model, all terms}
    \label{sm_fig:opinion_transitioning_model_all_terms}
    \end{center}
\end{figure}

Inspecting table \ref{sm_tab:experiment_vs_movement_regression} reveals that condition, phase one opinion, and confidence are all significant predictors of a participant transitioning opinion in the second phase. Participants in the CM condition have
half the odds of transitioning away from their promoted opinion, $O^1_1$ when compared with the C condition, the smaller odds of opinion changing for all opinions compared with C, since it reinforces the effects of confirmation bias in the C condition. Additionally, we see that with high levels of confidence the rate transitioning away decreases by $\frac{3}{4}$, as shown in Figure \ref{sm_fig:predicted_opinion_transition_frequency}. In this figure we highlight the promoted opinion for each condition with a yellow halo around the mean. Each effect described above can be seen, with corresponding increases in transitioning frequencies for opinions that are not promoted. Figure \ref{sm_fig:opinion_transitioning_model_all_terms} shows one plot for each parameter of the opinion transitioning model showing how each parameter is estimated to affect opinion transitioning frequency.

Table \ref{sm_tab:opinion_switching_anova_model_compare} shows the process of selecting the opinion transitioning model. As described in the main paper in Model Selection Section, we consecutively built increasingly complicated models by adding new parameters. We also had to decide how represent the baseline opinion rank predictor, with choices included a nominal, ordinal, or continuous values. The nominal value is the best performing because it does not enforce an ordered relation between the transitioning frequencies and the phase one opinions. This issue becomes apparent when examining the results for participants of the CS condition in Figure \ref{sm_fig:predicted_opinion_transition_frequency}. In this figure, the frequency of opinion changing for opinions $O_{1,3}^1$ increase, when compared to the C condition, while for opinions $O_2^1$, it decreases. If we were to try enforce a linear relationship between increasing opinion ranks, we would average away the divergent changing opinion behavior and miss the effect of external influence. Table \ref{sm_tab:opinion_switching_anova_model_compare} highlights the ANOVA comparison of each of the candidate models. Each increasingly complex model made a significant improvement over the previous model in the table.

\begin{table}[hp]
    \resizebox{1\textwidth}{!}{
    \begin{tabular}{l|ccccccccc}
        Model & \multicolumn{1}{l}{npar} & \multicolumn{1}{l}{AIC} & \multicolumn{1}{l}{BIC} & \multicolumn{1}{l}{logLik} & \multicolumn{1}{l}{Deviance} & \multicolumn{1}{l}{Chisq} & \multicolumn{1}{l}{$\Delta$Df} & \multicolumn{1}{l}{Pr(\textgreater{}Chisq)} & \multicolumn{1}{l}{Sig code} 
        \\ \hline
        Model.null & \ 2 & 3070.7 & 3082.2 & -1533.3 & 3066.7 & & & & \\
        4 Condition & \ 5 & 3048.7 & 3077.6 & -1519.3 & 3038.7 & \ 28.009 & \ 3 & 3.62E-06 & *** \\
        Inkblot + Opinion & \ 8 & 2863.9 & 2910.2 & -1424.0 & 2847.9 & 190.737 & \ 3 & 2.20E-16 & *** \\
        P1 Confidence & \ 9 & 2833.4 & 2885.5 & -1407.7 & 2815.4 & \ 32.546 & \ 1 & 1.16E-08 & *** \\
        P2 Confidence & 10 & 2818.6 & 2876.5 & -1399.3 & 2798.6 & \ 16.742 & \ 1 & 4.28E-05 & *** \\ \hline
        Condition x Rank (ordinal) & 13 & 2789.9 & 2865.2 & -1382.0 & 2763.9 & \ 34.685 & \ 3 & 1.42E-07 & *** \\
        Condition x Popularity & 13 & 2766.1 & 2841.4 & -1370.1 & 2740.1 & \ 58.500 & \ 3 & 1.22E-12 & *** \\
        Condition x Rank (nominal) & 21 & 2725.0 & 2846.6 & -1341.5 & 2683.0 & 115.670 & 11 & 2.20E-16 & *** \\
    \end{tabular}}
    \caption{\small{\textbf{ANOVA comparison for candidate transition frequency models.}
    This table compares the set of candidate models of transitioning frequency as a function of condition and other factors. The null model represents transitioning frequency purely as a random participant factor. Each model of the table is compared with the model of the previous row and the results of the likelihood ratio test are reported in the Chisq, $\Delta$Df and Pr(\textgreater{}Chisq) columns. The final three models of the table are not nested in each other, and instead represent the opinion ranks differently and their likelihood ratio test are performed against the P2 Confidence model. Using rank as a group or nominal model provides significant improvement over the model with ranks represented as ordinal numbers or popularity models even at a higher number of parameters.}}
    \label{sm_tab:opinion_switching_anova_model_compare}
\end{table}

\subsubsection{Promoted Opinion Model}

We can also model the effects of external influence by using the selection of the promoted opinion in the second phase as the response variable. Figure \ref{sm_tab:promoted_opinion_models} shows the results of three multinomial logistic regression models. Like the other regression models we use, we model users and inkblots are random effects. Each model predicts the selection of the promoted opinion for one of the influencing conditions CR, CM, and CS combined with C versus the C condition, with each model having its own column in the table. For the conditions combined with C condition, we label the promoted opinions for C condition and for the other influencing conditions. We do this to isolate the effects of reinforced confirmation bias and external influence from the confirmation bias signal present in all conditions. By comparing the C condition's coefficients with the other conditions we can isolate the effects of only confirmation bias. 


\begin{table}
\begin{center}
\begin{tabular}{l c c c}
\hline
 & CR Model & CM Model & CS Model \\
\hline
Condition (C), Rank 1 (Intercept) & $0.49$       & $1.62$       & $0.20^{**}$  \\
                          & $(1.51)$     & $(1.65)$     & $(1.77)$     \\
Inkblot                   & $1.07$       & $1.07$       & $0.95$       \\
                          & $(1.04)$     & $(1.06)$     & $(1.07)$     \\
$O_2^1$                   & $0.39^{***}$ & $0.14^{***}$ & $7.97^{***}$ \\
                          & $(1.18)$     & $(1.20)$     & $(1.20)$     \\
$O_3^1$                   & $0.22^{***}$ & $0.12^{***}$ & $1.25$       \\
                          & $(1.22)$     & $(1.23)$     & $(1.23)$     \\
Phase 1 Confidence        & $1.26^{*}$   & $0.96$       & $0.87$       \\
                          & $(1.09)$     & $(1.09)$     & $(1.10)$     \\
Phase 2 Confidence        & $1.41^{***}$ & $1.18$       & $1.10$       \\
                          & $(1.10)$     & $(1.09)$     & $(1.10)$     \\
Condition (CR)            & $1.63^{**}$  &              &              \\
                          & $(1.21)$     &              &              \\
Condition (CM)            &              & $2.16^{***}$ &              \\
                          &              & $(1.21)$     &              \\
Condition (CS)            &              &              & $4.01^{***}$ \\
                          &              &              & $(1.22)$     \\
\hline
AIC                       & $1342.24$    & $1335.59$    & $1278.81$    \\
BIC                       & $1388.13$    & $1381.55$    & $1324.70$    \\
Log Likelihood            & $-662.12$    & $-658.80$    & $-630.40$    \\
Num. obs.                 & $1210$       & $1220$       & $1210$       \\
Num. groups: user         & $121$        & $122$        & $121$        \\
Num. groups: inkblot      & $10$         & $10$         & $10$         \\
Var: user (Intercept)     & $0.46$       & $0.48$       & $0.48$       \\
Var: inkblot (Intercept)  & $0.05$       & $0.22$       & $0.35$       \\
\hline
\multicolumn{4}{l}{\scriptsize{$^{***}p<0.001$; $^{**}p<0.01$; $^{*}p<0.05$}}
\end{tabular}
\caption{\small{\textbf{Promoted opinion frequency regression models}. 
        	This table defines the parameters and coefficients of our mixed effects model of promoted opinion selection. There are three such models, one for each pair of C/R, C/M, and C/S. The DV for this model is 0 if the participant didn't select the promoted opinion in the second phase and 1 if they did. Since each condition has a different promoted opinion(s) the C condition measures its 'promoted' selection frequency using the same opinion ranking as the condition it is modeled with, $O_1^1$ for M, $O_2^1$ for S, and $O_i^1 \to $$O_i^2$ for R.} Each condition is significant in its model, demonstrating that the condition is a significant indicator for selecting the promoted opinion. The beta coefficients, when converted to odds ratios, show a 1.63x increase in odds of selecting promoted for R, 2.16x odds for M, and 4.01x odds for CS when compared to C.}
    \label{sm_tab:promoted_opinion_models}
\end{center}
\end{table}

\subsubsection{The basic analyses of phases one and two}
This section outlines basic information about the opinion selection data in both phases. Here we will outline the distributions of opinion selection per phase and condition, outline basic opinion flow information, and show supporting tables and figures to supplement the main text. 
 When analyzing the second phase rankings, we add a hat over the $O$, like so: $\hat{O_r^p}$.

\begin{table*}[h!]
    \centering
    \begin{adjustbox}{width=0.85\columnwidth,center}
    \begin{tabular}{llllllllll}
    \textbf{rank}      & \multicolumn{3}{c}{$O_1^p$} \textbf{(most popular)}   & \multicolumn{3}{c}{$O_2^p$}  & \multicolumn{3}{c}{$O_3^p$} \textbf{(unpopular)}                 \\
    \textbf{phase}     & \textbf{P1} & \textbf{P2} & \textbf{Change} & \textbf{P1} & \textbf{P2} & \textbf{Change} & \textbf{P1} & \textbf{P2} & \textbf{Change} \\
    \textbf{Control}   & 0.59        & 0.57        & -0.0163         & 0.23        & 0.24        & 0.0065          & 0.17& 0.18& 0.009\\
    \textbf{Recall}      & 0.58        & 0.59        & 0.0133          & 0.27        & 0.24        & -0.0333         & 0.14& 0.16& 0.02\\
    \textbf{Most-popular}  & 0.57        & 0.7         & \textbf{0.1278}          & 0.26        & 0.18        & \textbf{-0.0786}         & 0.17& 0.12& -0.049\\
    \textbf{Second-most-popular} & 0.58        & 0.4         & \textbf{-0.1866}         & 0.25        & 0.48        & \textbf{0.2316 }         & 0.16& 0.12& -0.045\end{tabular}
    \end{adjustbox}
    \caption{\textbf{Opinion composition per condition per phase}. This table shows for each opinion the selection frequencies in both phases, as well as the change between phases. The bold entries show the first signs of interesting behavior. The participants in the CM condition increase the average fraction of opinions in $O_1^2$, the promoted opinion of that condition. This same behavior is shown by participants in the CS condition, where CS's promoted opinion, $O_2^2$. increases the selection frequencies when compared to those from the first phase. It is important to note that the P1 column for each condition is different from each other because the fraction is computed from the group of each condition independently for this table, and for each condition there is slight variation in the fraction of participants who chose each opinion.}
    \label{sm_tab:opinion_population_change}
\end{table*}

\subsubsection{Opinion transition matrices}
\begin{figure}[hp]
    \begin{center}
    \begin{subfigure}[t]{0.4\textwidth}
    \centering
        \includegraphics[width=1\linewidth]{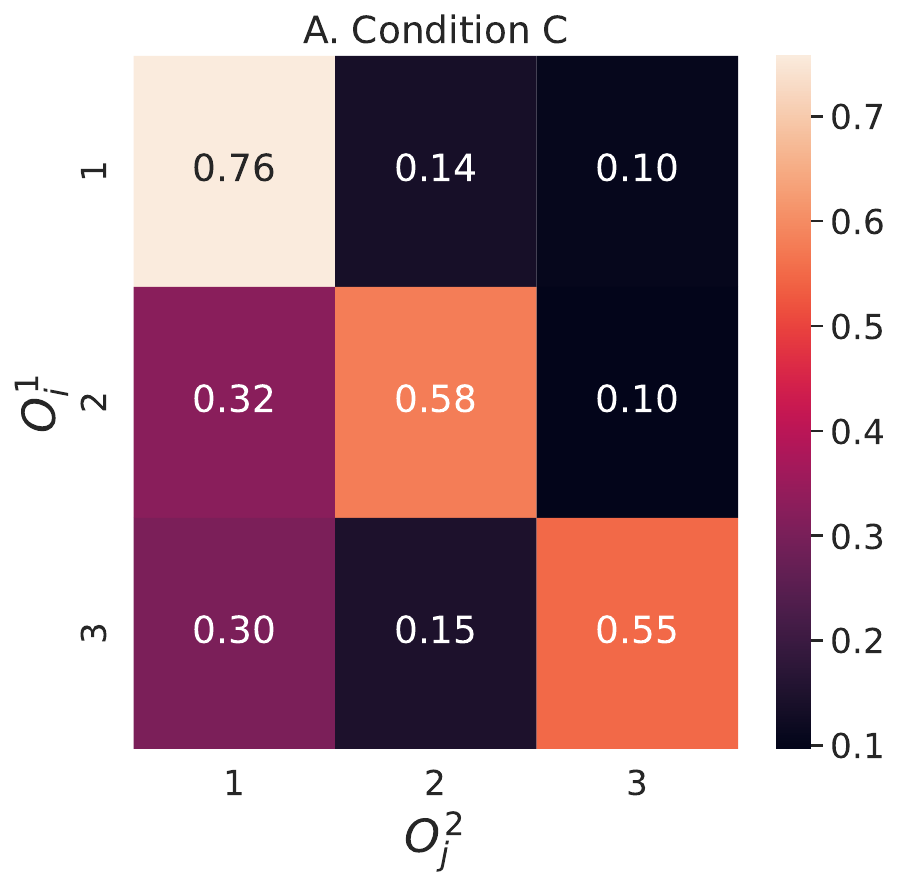}
    \end{subfigure}%
    \centering
    \begin{subfigure}[t]{0.4\textwidth}
    \centering
        \includegraphics[width=1\linewidth]{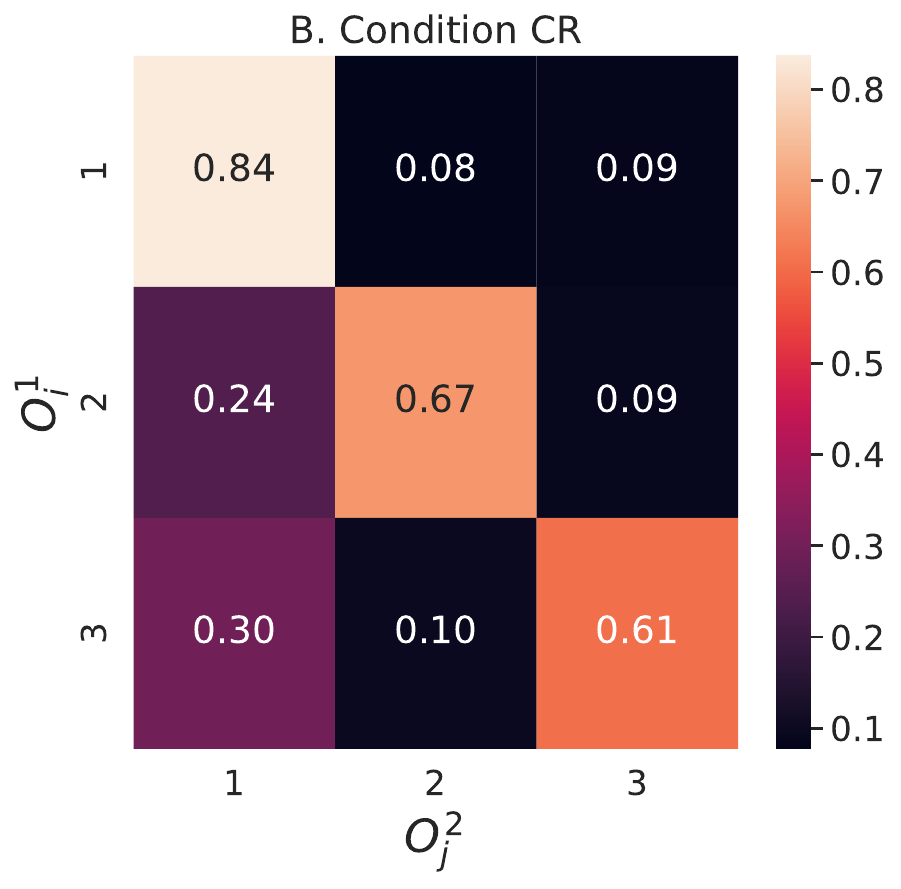}
    \end{subfigure}
    
    \begin{subfigure}[t]{0.4\textwidth}
    \centering
        \includegraphics[width=1\linewidth]{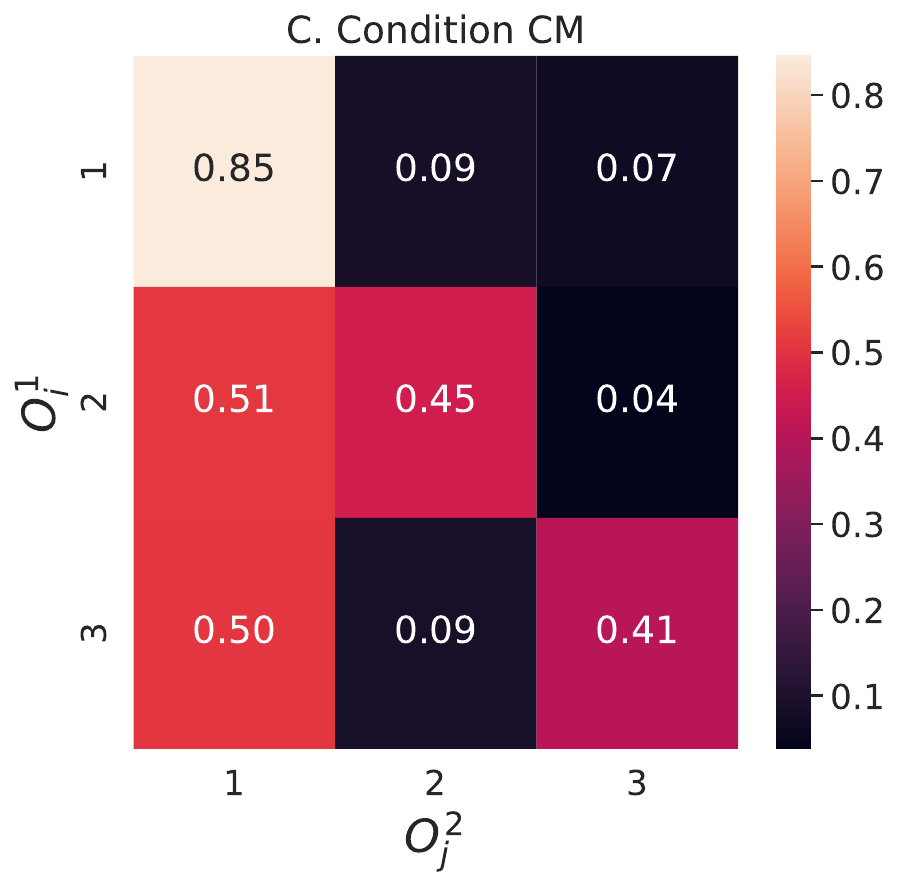}
    \end{subfigure}%
    \centering
    \begin{subfigure}[t]{0.4\textwidth}
    \centering
        \includegraphics[width=1\linewidth]{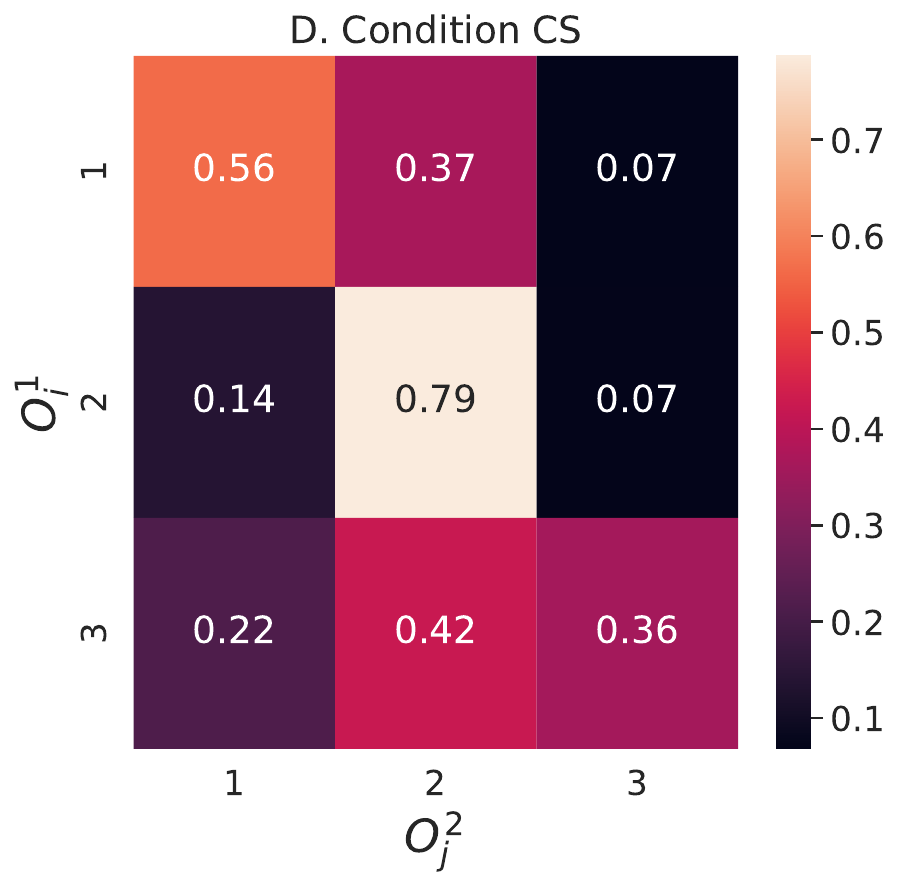}
    \end{subfigure}
    \caption{\textbf{Opinion transition matrix}. This figure defines the fractional flow of opinions, $O_i^1 \to O_j^2$ for some opinion $i$ in the first phase to some opinion $j$ in the second phase, for each condition.}
    \label{sm_fig:p1_p2_transition}
    \end{center}
\end{figure}

\begin{figure}[hp]
\centering

\includegraphics[scale=0.38]{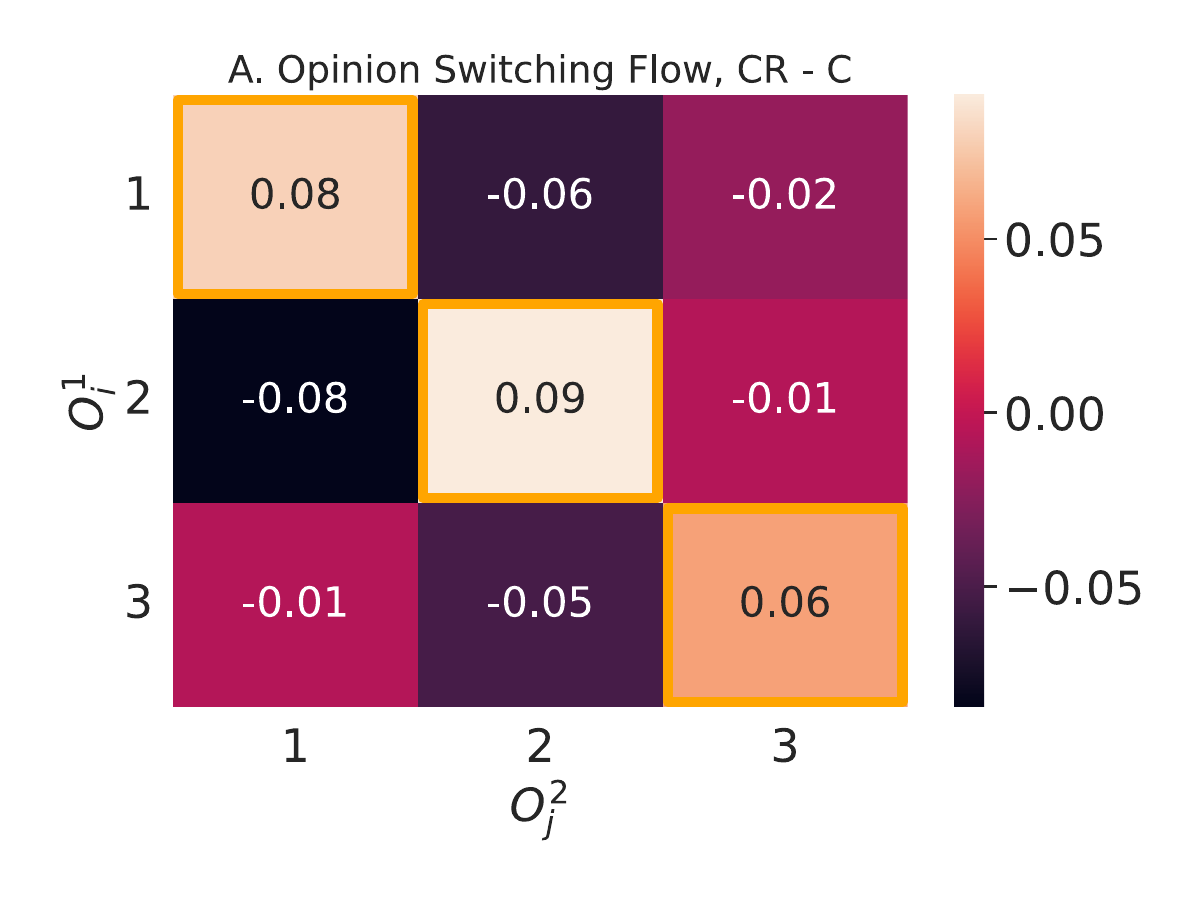}%
\includegraphics[scale=0.38]{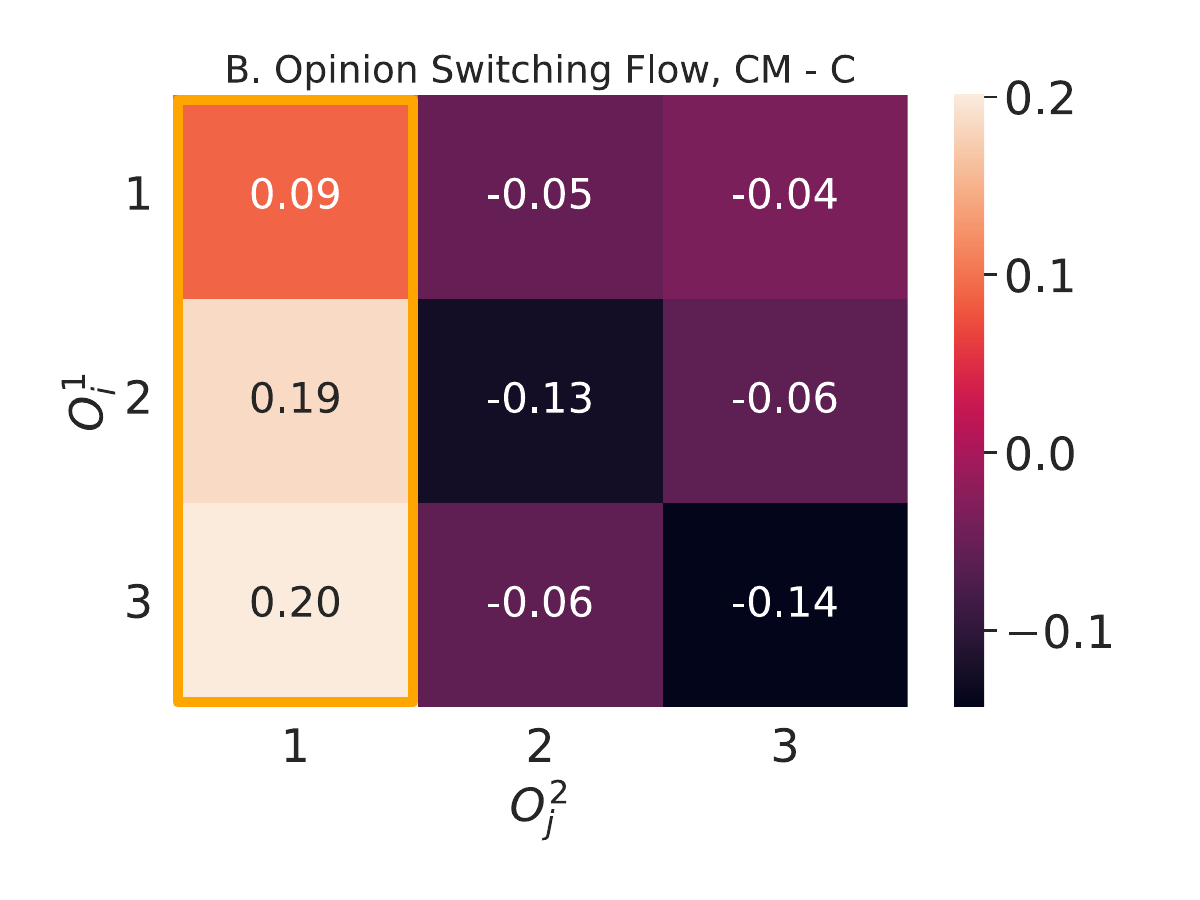}
\includegraphics[scale=0.4]{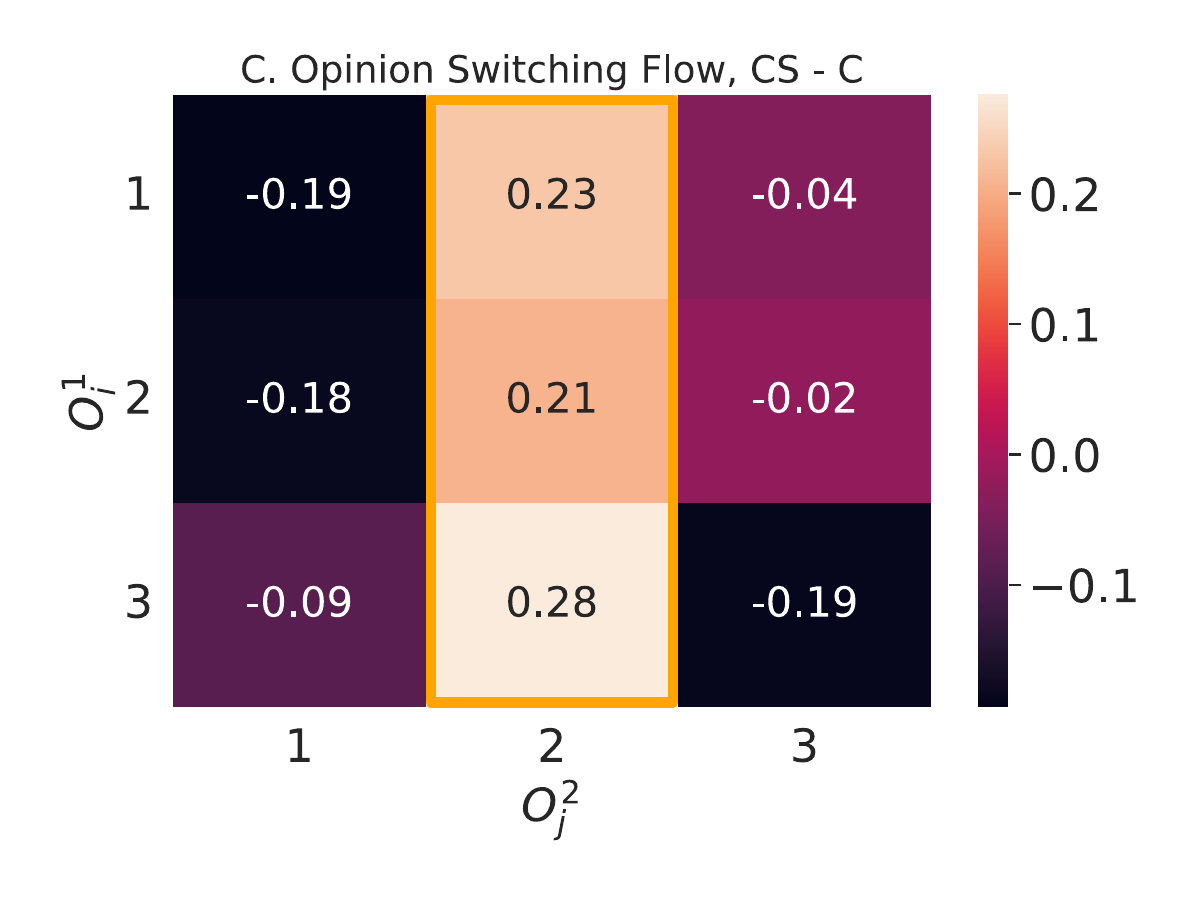}
\caption{\small{\textbf{Opinion transition matrices for CR, CM, and CS conditions compared with C.} This figure defines the fractional flow of opinions, $O_i^1 \to O_j^2$ for some opinion i in the first phase to some opinion j in the second phase, relative to such frequencies from the C condition}. A positive value indicates a greater frequency versus control while a negative value indicates the opposite. The highlighted column in each subplot (bounded in yellow) refers to the promoted opinion of that condition.}
\label{sm_fig:full_fig_move_diff}
\end{figure}

Here, we visualize how nine transitions can be embedded in a $3\times3$ transition matrix T, see Figure \ref{sm_fig:p1_p2_transition}. Let $f(O_{1,2,3}^{X,p})$ denote the fraction of participants in the current condition $X$ and the opinion $O_r^p$. Then, each element of matrix $T$ is $T[O_{1,2,3}^1,O_{1,2,3}^2]=f(O_r^2)-f(O_s^2)$. 

Figure \ref{sm_fig:full_fig_move_diff} compares the participants' transition frequencies in CR, CM, and CS with those in the C condition. These matrices visualize clearly distinct participant opinion transition patterns in the different influence conditions. In the CM and CS  conditions, a single column corresponding to the rank of the promoted opinion contains all positive values. These columns represent a higher transition rates of participants than in the Confirmation bias condition. In the CM and CS conditions, the rate at which participants selected the promoted opinions, {$O_1^2$} and increased by $\sim$17\%, and $\sim$24\%, respectively compared to the Confirmation bias condition. Still, the number of participants added to the promoted opinion in the CM condition is larger than that in the CS condition.

\subsubsection{Promoted opinion analysis}
\begin{figure}[b!]
    \centering
    \includegraphics[width=.95\linewidth]{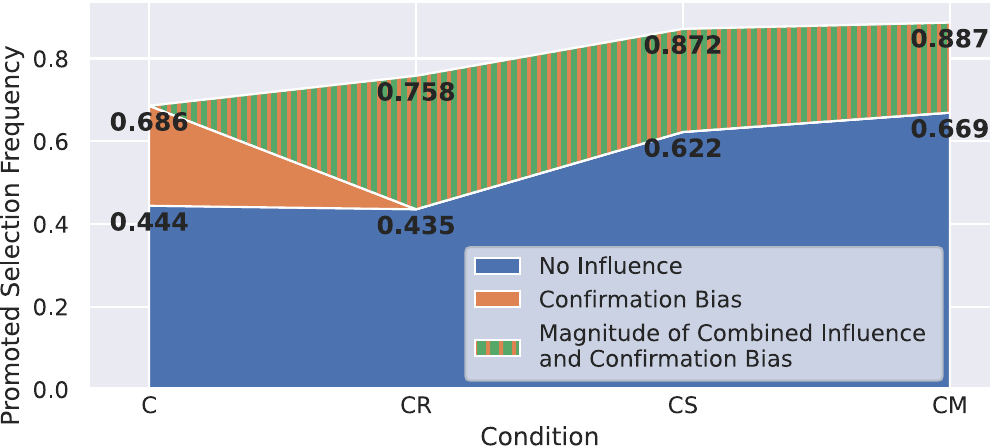}
    \caption{\small\textbf{The magnitude of influence on promoted opinion selection.
    The figure shows the cumulative components in selection of the promoted opinion over different conditions. Values computed in CR, CM and CS include both the promoted opinion confirmation bias and R, M, and S influences, respectively.}}
    \label{fig:influence_v_opinion}  
\end{figure}

For our second method for verifying our hypothesis, we used the data collected in our experiment to compute fractions of populations that are P members in each condition. Then, in each condition, we measured the distribution of selections of the promoted opinion under no influence in the first phase and denote as $P^1_x$ fraction of participants in condition $x\in\{C,CR,CM,CS\}$ with promoted opinion. As shown in Fig.~\ref{fig:influence_v_opinion}, in C $P^1_C=0.444, P^1_{CR}=0.435$, $P^1_{CM}=0.669$, and $P^1_{CS}=0.622$. The large difference between them are caused by popularity of promoted opinions which are the most popular opinion in CM, the second most popular one in CS, but any opinion in C and CR. That means that for the participants in C only $1 - 0.444 = 0.556$ and in CR $0.565$ of the populations can change their opinions to the promoted one, while for the participants in CM and CS these fractions are just 0.331 and 0.378, respectively. To measure the strength of influence of the promoted opinion in condition $x$, we use the formula: $(P^2_x-P^1_x)/(1-P^1_x)$ which represent the fraction of participants not in $P^1_x$ in condition $x$ joint $P^2_x$. The two highest such fractions are CM at 0.658 and CS at 0.660, while lower fractions are 0.572, and 0.436 for CR and C, respectively. Comparing these measures to results of converge regression odds shows that the decreasing order of conditions stays the same for both methods.   

\subsection{Exploratory Analyses}
\label{sm_section:exploratory_analysis}
Here, we provide additional information on the exploratory analyses presented in the main text.

\subsubsection{\textcolor{black}{Effects of confidence on the impact of ambiguous influence}}

To determine the effects of confidence on the rate of conforming with or opposing anonymous influence we created the following Bayes multi-level regression model: 

$conformed \sim confidence_{phase 1} \times O^1 \times condition + (1|user) + (1|inkblot)$. 

This model predicts conformity as a binary variable that is 1 if the participant conforms with (selects) the promoted opinion presented to them in phase two of the study. The predictors are the confidence level (1-4) the participant has in their phase 1 opinion, the rank of the opinion held in phase 1 ($O^1$), the condition the participant was apart of in phase 2, and the interaction between these terms. With this model, we can see how frequently participants of different confidence levels conform to ambiguous influence and how this differs for participants whose opinion was the same or different to this promoted opinion in phase 1. We summarize the results of this model in table \ref{sm_tab:confidence_vs_conformity}. The intercept for this model is the lowest confidence level for participants whose phase 1 opinion was rank 1, ($O_1^1$), for CM condition participants. The coefficient for this intercept is the odds that a participant of the CM condition who held opinion rank 1 in phase 1 and was not at all confident in their response would side with the promoted opinion in phase 2. The rest of the coefficients are odds ratios against this odds, so the value of $1.55$ for P1 Confidence (Linear) shows us that as the confidence levels go up for CM participants with $O^1_1$ we increase the odds of selecting the promoted opinion in phase 2 by $1.55$ times. However, this relationship is not significant due to the Credible Interval (CI) overlapping 1, so the odds ratio, versus the reference level, could possibly be 1 within 95\% CI.

The main points of interest in this table are the P1 Confidence (Linear), $O^1_{1-3}$, Condition (CS) and their interactions. Since confidence is an ordered factor, three polynomial contrasts were fit against the confidence data: linear, quadratic, and cubic, but only linear coefficients are shown in this table due to the insignificance of the other terms and to keep the table more readable.

\begin{table}[h!]
	\Large
	\centering
    \resizebox{.9\textwidth}{!}{
	\begin{tabular}{|l r r|}
	\hline
    	Predictors & Odds Ratios & CI \\
    	\hline
    	Intercept (Not at all confident, $O^1_1$, CM)& 7.83 & 4.72 - 13.29 \\
    	P1 Confidence (Linear) & 1.55 & 0.70 - 3.50 \\
            $O^1_2$ & 0.14 & 0.08 - 0.25 \\
            $O^1_3$ & 0.13 & 0.07 - 0.23 \\
            Condition (CS) & 0.07 & 0.04 - 0.13 \\
            P1 Confidence (Linear) x $O^1_2$ & 0.12 & 0.03 - 0.40 \\
            P1 Confidence (Linear) x $O^1_3$ & 0.41 & 0.10 - 1.57 \\
            P1 Confidence (Linear) x Condition (CS) & 0.28 & 0.10 - 0.75 \\
            $O^1_2$ x Condition (CS) & 73.90 & 32.41 - 178.10 \\
            $O^1_3$ x Condition (CS) & 10.16 & 4.40 - 23.43 \\
            P1 Confidence (Linear) x $O^1_2$ x Condition (CS) & 74.71 & 11.17 - 534.96 \\
            P1 Confidence (Linear) x $O^1_3$ x Condition (CS) & 7.25 & 1.09 - 48.60 \\
     \hline
     \end{tabular}
     \begin{tabular}{|l|l|}
     \hline
        \multicolumn{2}{|c|}{Random Effects} \\
    	\hline
        $\sigma^2$ & 3.29 \\
        $\tau_{00\ user}$ & 0.15\\
        $\tau_{00\ inkblot}$ & 0.88\\
        ICC & 0.24 \\
        $N_{user}$ & 121 \\
        $N_{inkblot}$ & 10 \\
        \hline
        \hline
    	Numerical Observations & 1210 \\
    	Marginal $R^2$ & 0.239\\
    	Conditional $R^2$ & 0.316\\
	\hline
	\end{tabular}}
  \caption{\small{\textbf{Multilevel logistic regression of opinion transitioning frequencies.}}}
  \label{sm_tab:confidence_vs_conformity}
\end{table}

We next perform hypothesis tests to answer the following four questions (for convenience we will rename P1 Confidence (Linear) to Conf and Condition (CS) to CS, to make things more readable):
\begin{itemize}
    \item $Q_1$: For the $O_1^1$ opinion of the CM condition, does confidence have a significantly positive or negative impact on conformity?
    \begin{itemize}
        \item Hypothesis $H_1$: Conf $\neq$ 0
    \end{itemize}

    \item $Q_2$: For the $O_2^1$ opinion of the CM condition, does confidence have a significantly positive or negative impact on conformity?
    \begin{itemize}
        \item Hypothesis $H_2$: Conf + Conf:$O^1_2$ $\neq$ 0 
    \end{itemize}

    \item $Q_3$: For the $O_1^1$ opinion of the CS condition, does confidence have a significantly positive or negative impact on conformity?
    \begin{itemize}
        \item Hypothesis $H_3$: Conf + Conf:CS  $\neq$ 0 
    \end{itemize}

    \item $Q_4$: For the $O_2^1$ opinion of the CS condition, does confidence have a significantly positive or negative impact on conformity?
    \begin{itemize}
        \item Hypothesis $H_4$: Conf + Conf:CS + Conf:$O^1_2$ + Conf:$O^1_2$:CS  $\neq$ 0 
    \end{itemize}
    
\end{itemize}

To do this, we utilize the hypothesis method of the brms package in r. The results of which are shown in table \ref{sm_tab:confidence_conformity_hypotheses}.

\begin{table}[h]
\centering
\begin{adjustbox}{center}
\begin{tabular}{|rrrrr|}
  \hline
  Hypothesis & Estimate (log Odds Ratio) & Est.Error & CI (2.5\%) & CI (97.5\%)  \\
  \hline
    *$H_1$ & -1.72 & 0.53 & -2.81 & -0.69 \\
    *$H_2$ & -0.84 & 0.33 & -1.51 & -0.20 \\
    $H_3$ & 0.44 & 0.41 & -0.36 & 1.25 \\
    *$H_4$ & 1.31 & 0.68 & 0.07 & 2.77 \\
   \hline
\end{tabular}
 \end{adjustbox}
 \caption{\textbf{Hypothesis outcomes for confidence vs conformity} This table shows the results of the four hypothesis tests evaluating how confidence effects conforming to the promoted opinion for the CM/CS conditions and $O^1_1$ / $O^1_2$. We use * to indicate hypotheses that are significant, which are hypotheses where the log odds ratio of 0 lies outside the 95\% CI of the Posterior.}

 One interesting behavior seen in this confidence analysis is the lack of correlation of confidence and opinion switching for fringe / unpopular opinions ($O_3^1$). As can be seen in Table \ref{sm_tab:confidence_vs_conformity} the relationship between confidence and $O_3^1$ has a very wide CI that is both well below and well above 1, indicating that the relationship could be decreasing increasing or nothing within $95\%$ confidence. This result is different to the behaviors we see in both $O_1^1$ and $O_2^1$. Looking at the raw response counts we see that $O_3^1$ has $3/5$ as many datapoints as $O_2^1$ and $1/4$ as many datapoints as $O_1^1$. This decrease in datapoints within $O_3^1$ is enough to make it so that the model is unable to distinguish the relationship between confidence and selecting the promoted opinion for this group.

 \label{sm_tab:confidence_conformity_hypotheses}
\end{table}

\pagebreak

\section{Figures}

 \begin{figure}[h!]
    \centering
        \includegraphics[width=.95\textwidth]{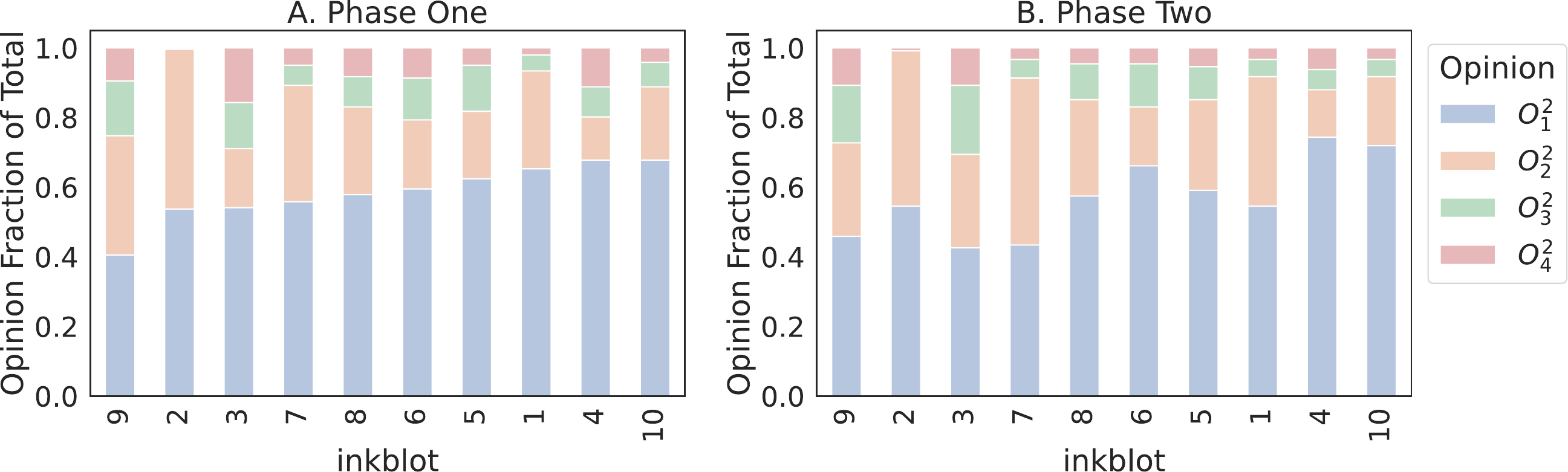}
    \caption{\textbf{Inkblot opinion populations for phases one and two.} This figure shows the overall opinion selection distribution for each inkblot for the first and second phases of the experiment. The bars are made up of up to four colors, one for each opinion $O_1^1 ... O_4^1$. Each color's size is proportional to the number of votes the opinion had in the given phase. The inkblots are ordered by increasing size of the opinion in the first phase.}
    \label{sm_fig:p1_opinion_groups_msorted}
\end{figure}

\begin{figure}[hp]

    \includegraphics[width=.95\textwidth]{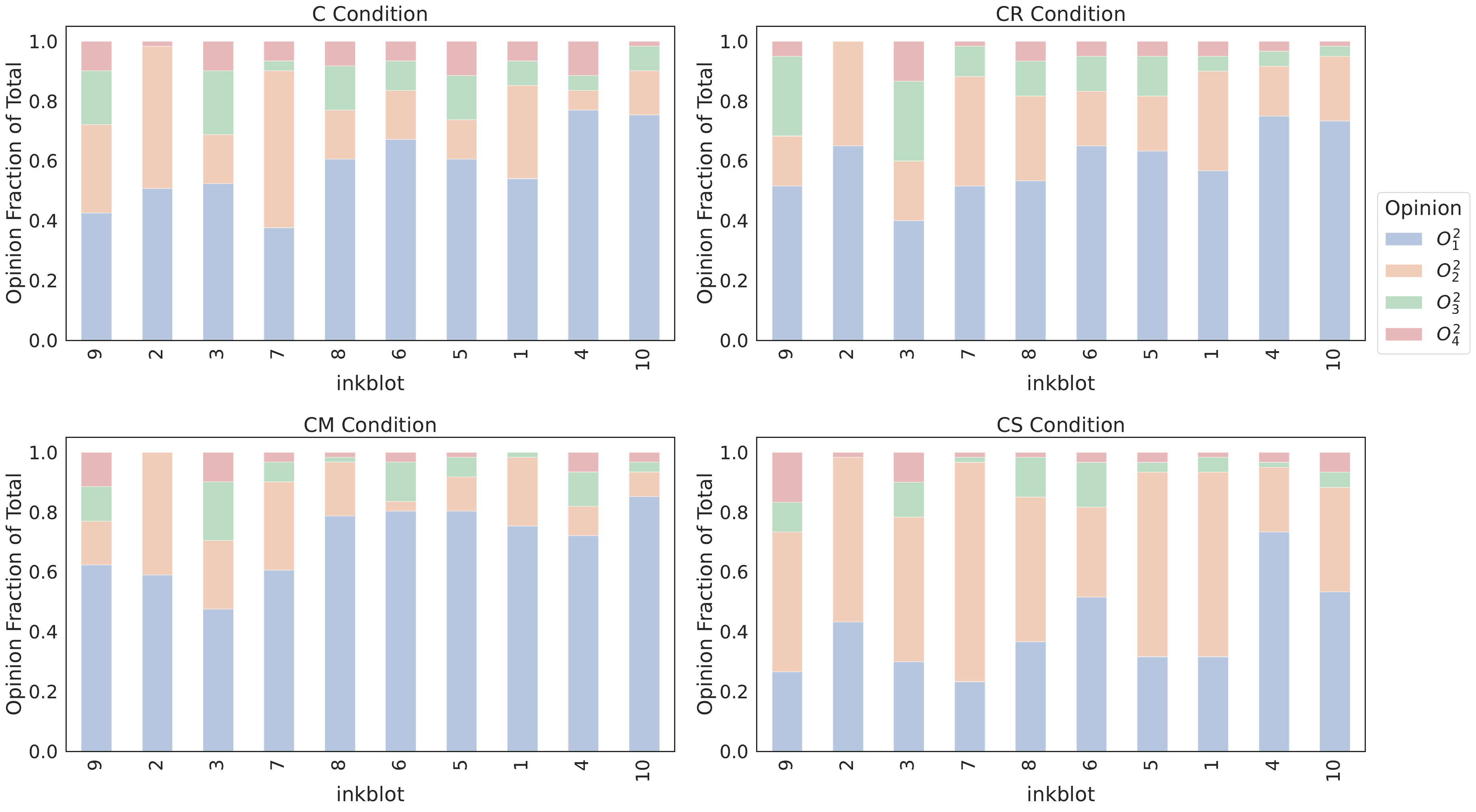}
    \caption{\textbf{Phase 2 inkblot opinion populations, all conditions}. This figure shows, for each condition and inkblot, the opinion selection distribution for that condition and inkblot. The bars are made up of up to four colors, one for each opinion $O_1^1 ... O_4^1$. Each color's size is proportional to the number of votes the opinion had in the second phase. The inkblots are ordered by increasing size of the opinion in the first phase.}
\end{figure}

\begin{figure}[hp]
    \begin{center}
    \begin{subfigure}[t]{0.5\textwidth}
    \centering
        \includegraphics[width=.5\linewidth]{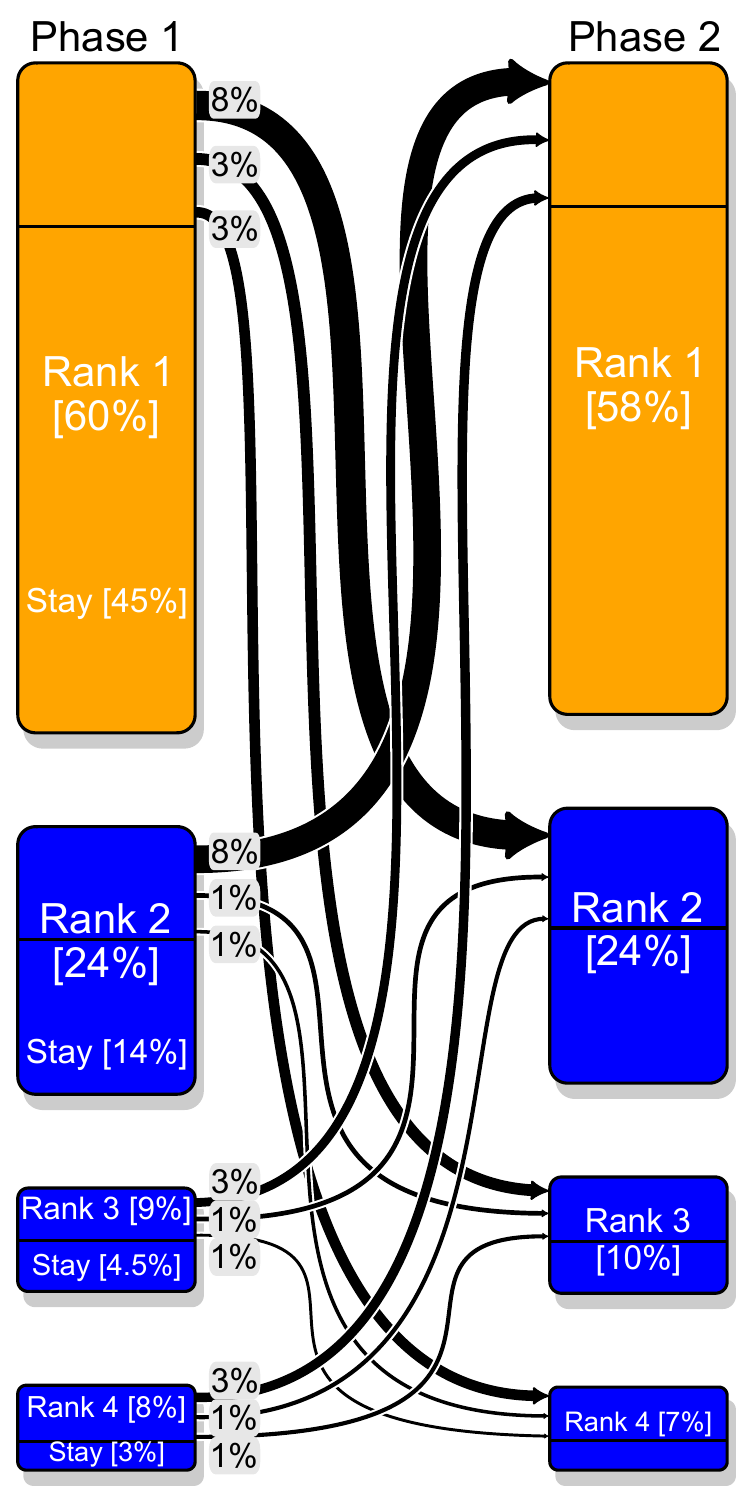}
    \caption{Control}
    \label{sm_fig:C_opinion_transition}
    \end{subfigure}%
    \centering
    \begin{subfigure}[t]{0.5\textwidth}
    \centering
        \includegraphics[width=.5\linewidth]{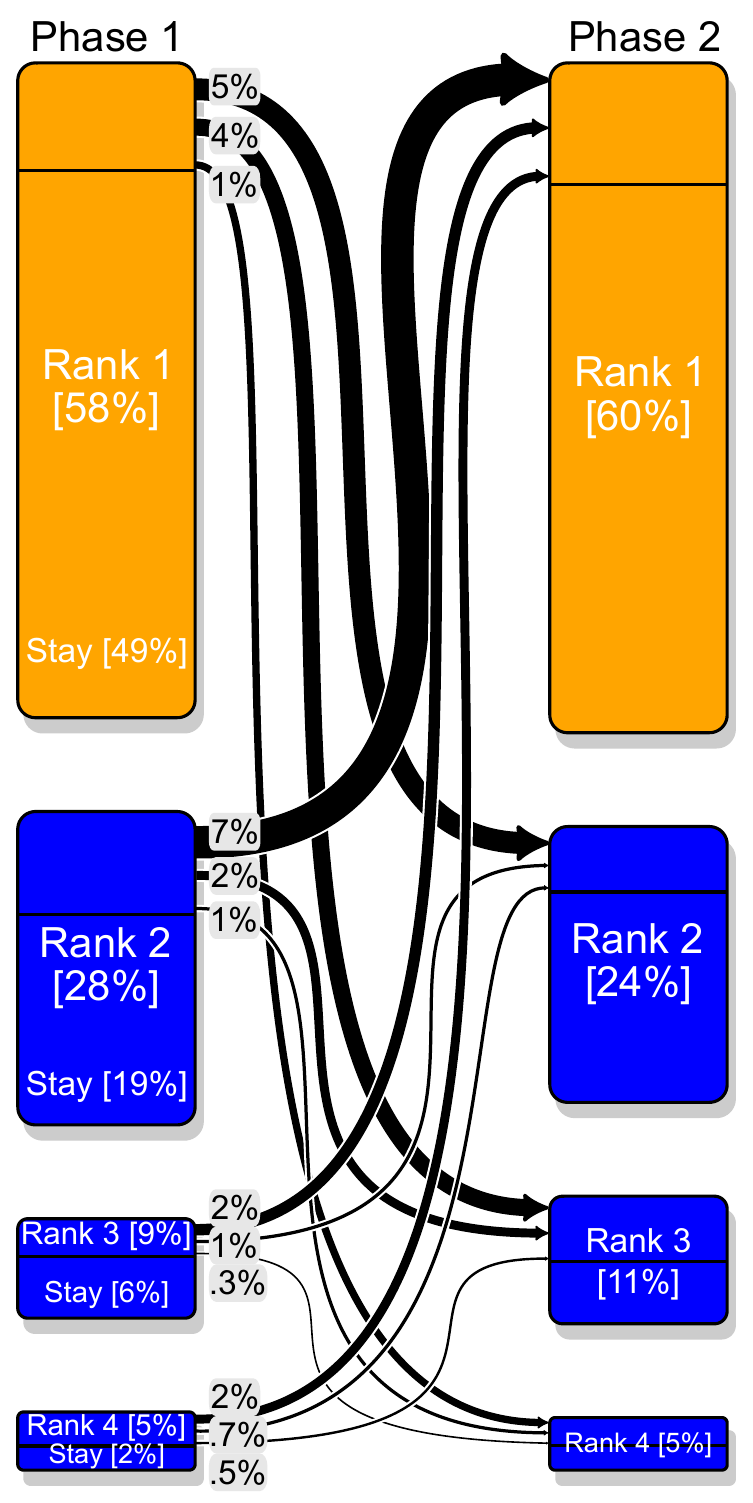}
    \caption{Recall}
    \label{sm_fig:R_opinion_transition}
    \end{subfigure}
    \begin{subfigure}[t]{0.5\textwidth}
    \centering
        \includegraphics[width=.5\linewidth]{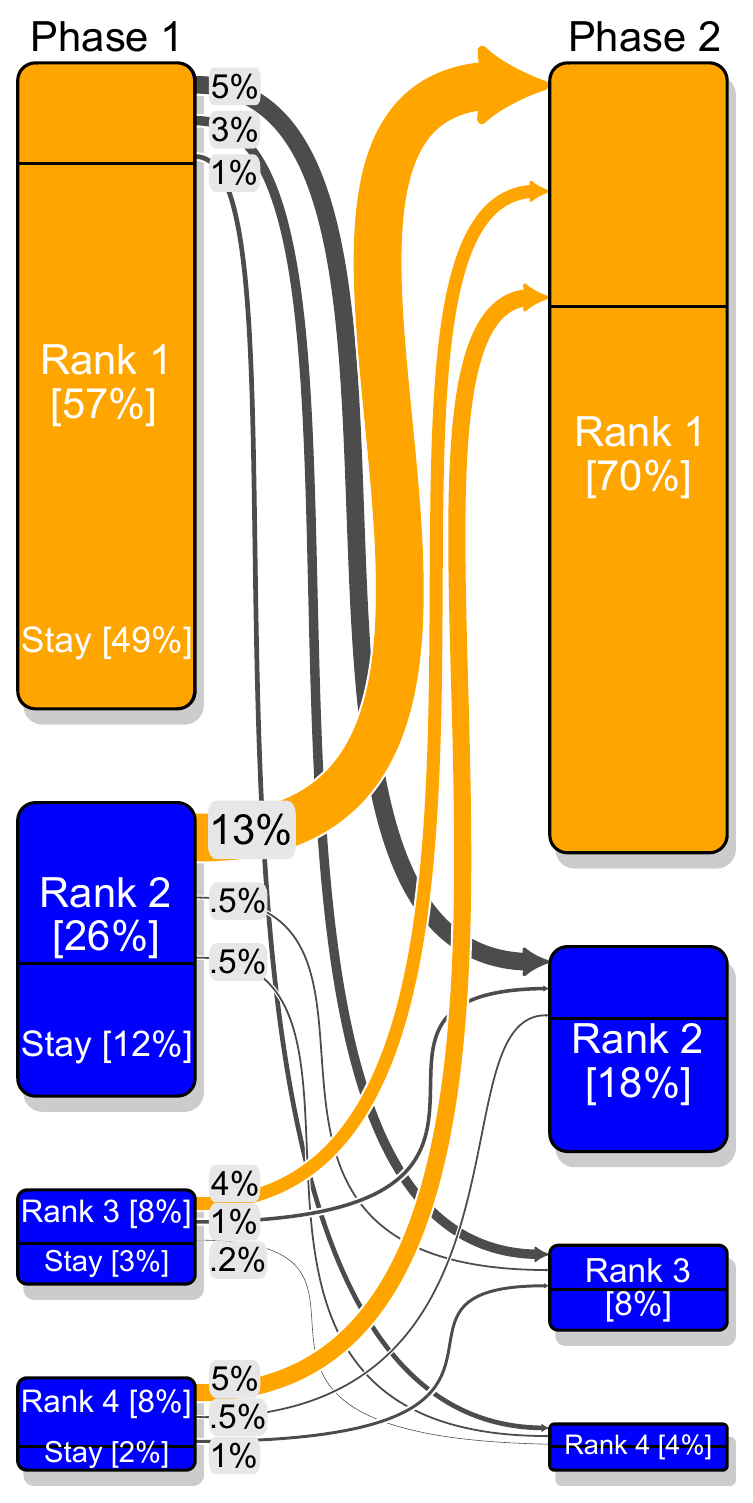}
    \caption{Most-Popular}
    \label{sm_fig:M_opinion_transition}
    \end{subfigure}%
    \centering
    \begin{subfigure}[t]{0.5\textwidth}
    \centering
        \includegraphics[width=.5\linewidth]{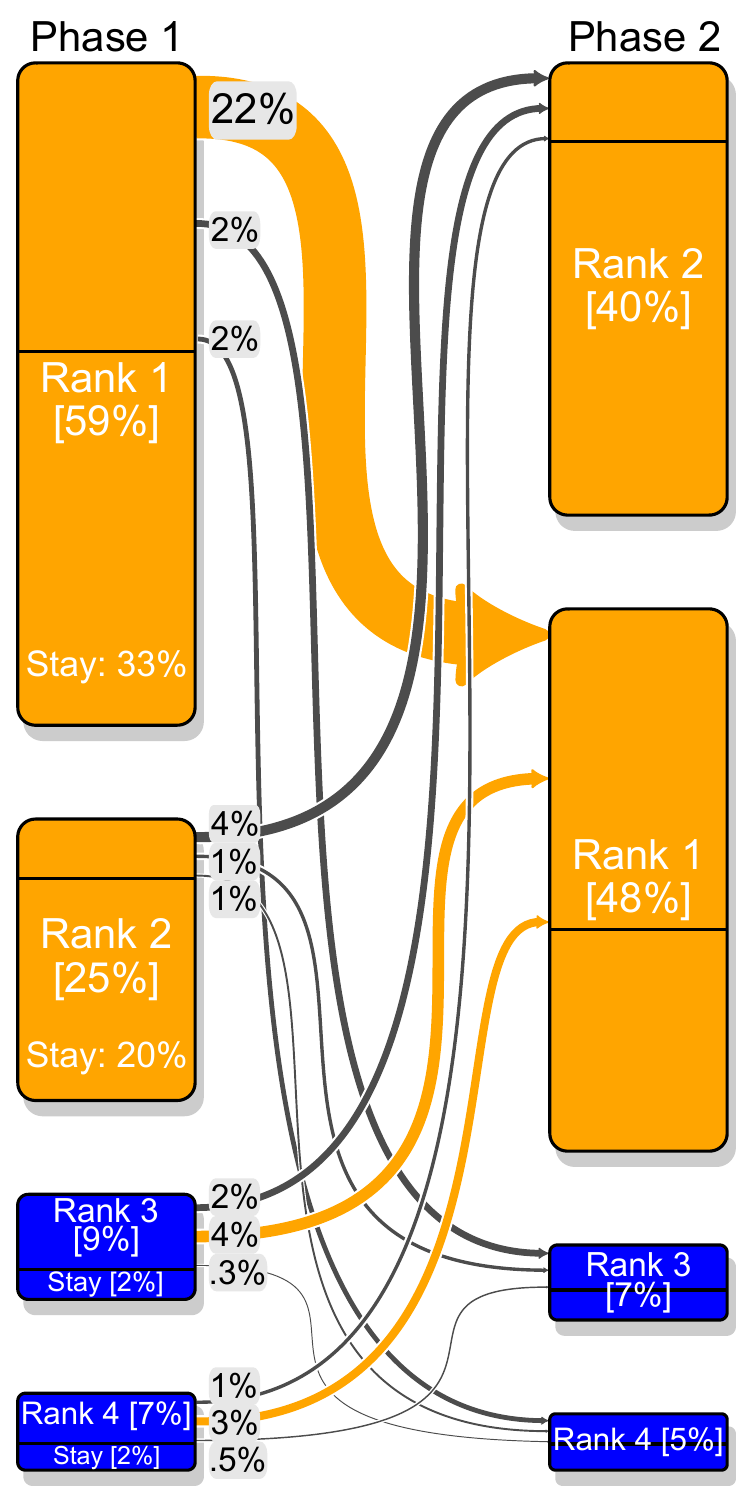}
    \caption{Second-Most-Popular}
    \label{sm_fig:S_opinion_transition}
    \end{subfigure}
    \caption{\textbf{Opinion Flow by Experimental Condition}. This figure shows a sankey diagram representing, visually, the flow of opinions from phase one to phase two ($O_i^1 \to O_j^2$). The order of the boxes in each column is based on the first phase rank. We can see that for the (d) plot that $O_2^1$ becomes $\hat{O}_1^2$, meaning ambiguous influence was able to make the lower ranked opinion popular enough to overcome the most popular opinion from the first phase.}
    \label{sm_fig:opinion_transitions_flowchart}
    \end{center}
\end{figure}

\begin{figure}
    \centering
        \includegraphics[height=.85\textheight]{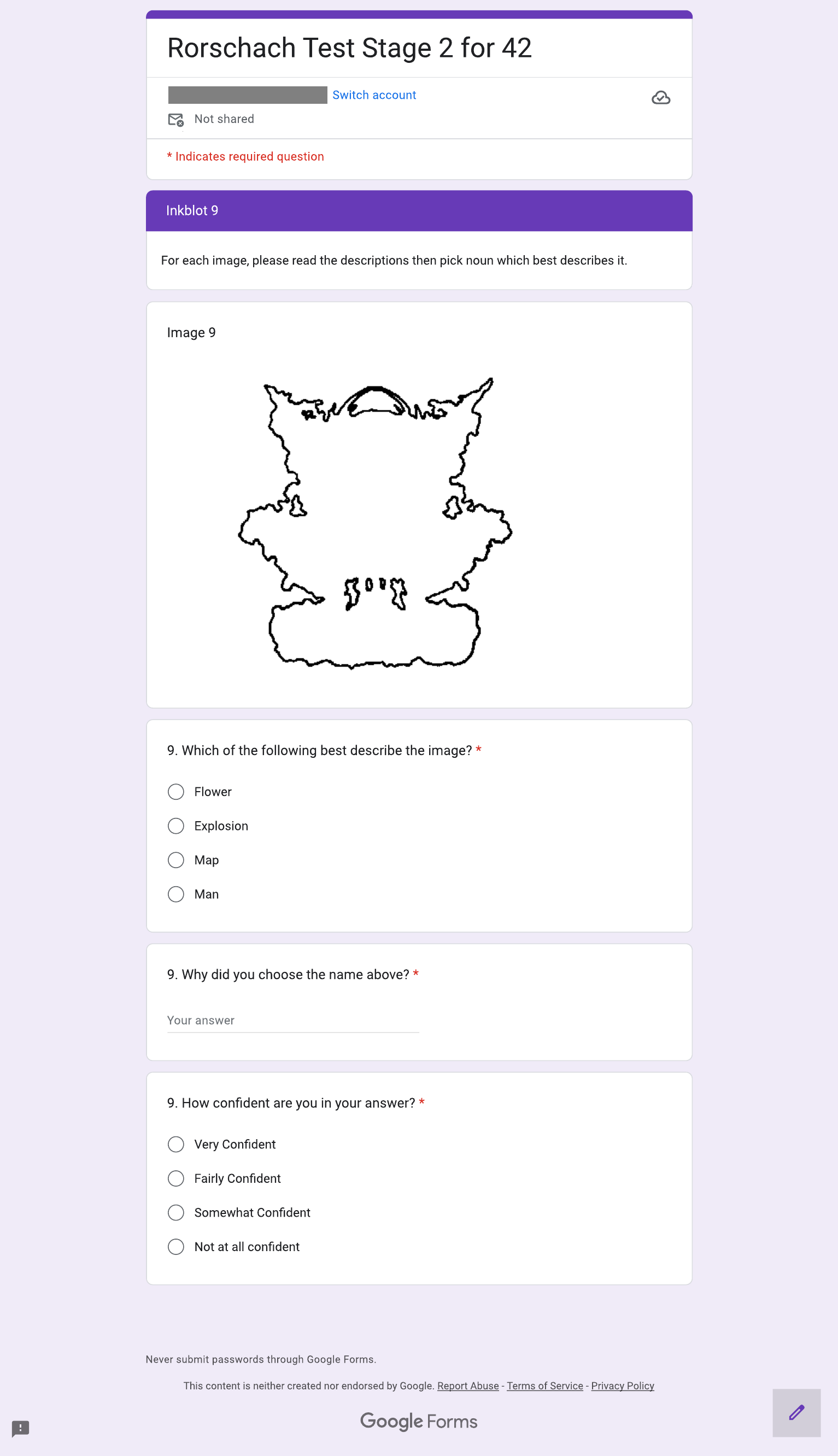}
    \caption{\textbf{Example survey for the C condition, inkblot 9} Here we show what a survey would have looked like for a participant of the C condition in our study. While only one is shown for brevity, each inkblot had its own page with the same information presented.}
    \label{sm_fig:example_survey_C_condition}
\end{figure}

\begin{figure}
    \centering
        \includegraphics[height=.85\textheight]{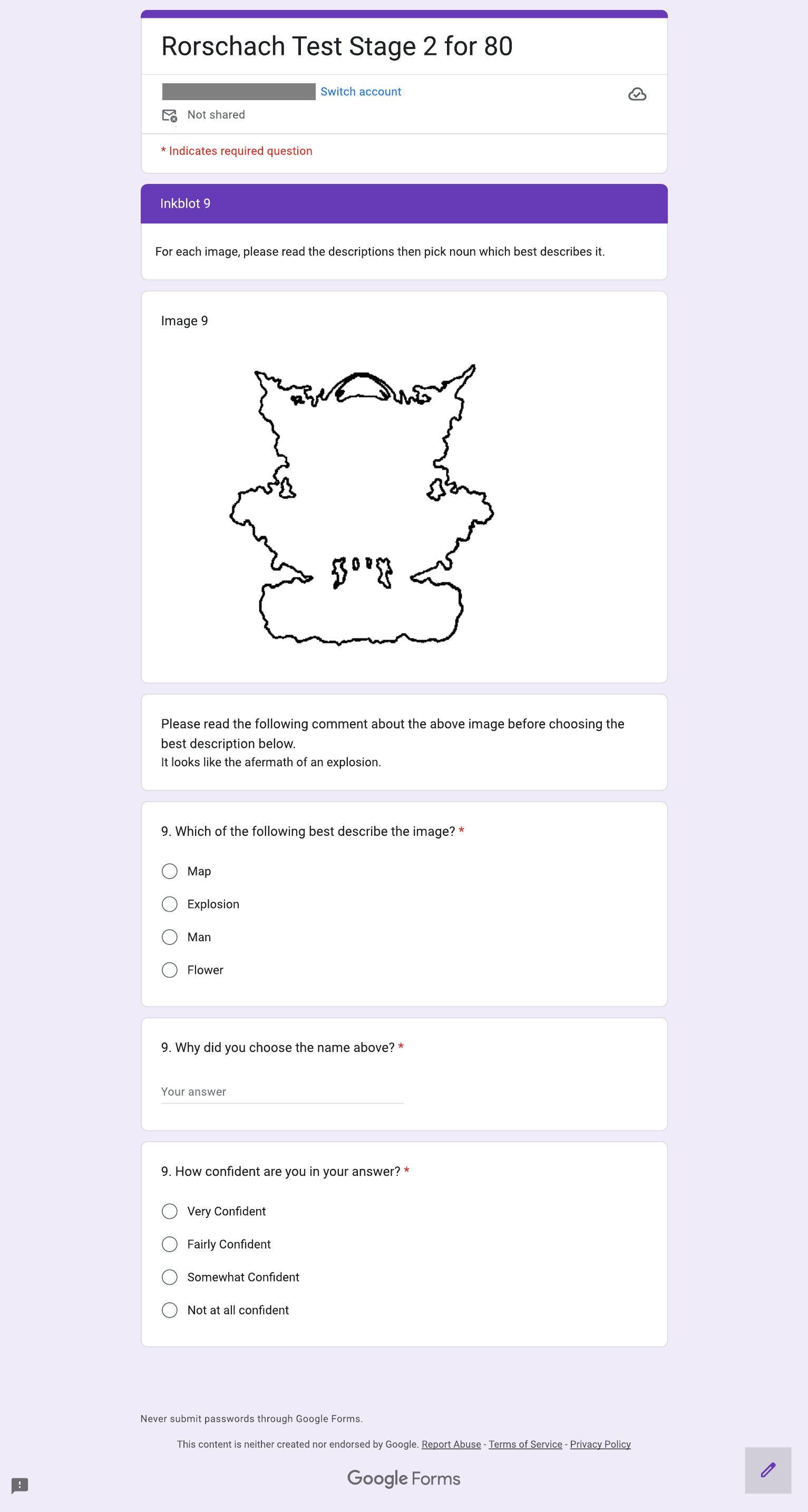}
    \caption{\textbf{Example survey for the CR condition, inkblot 9} Here we show what a survey would have looked like for a participant of the CR condition in our study. While only one is shown for brevity, each inkblot had its own page with the same information presented.}
    \label{sm_fig:example_survey_CR_condition}
\end{figure}

\begin{figure}
    \centering
        \includegraphics[height=.85\textheight]{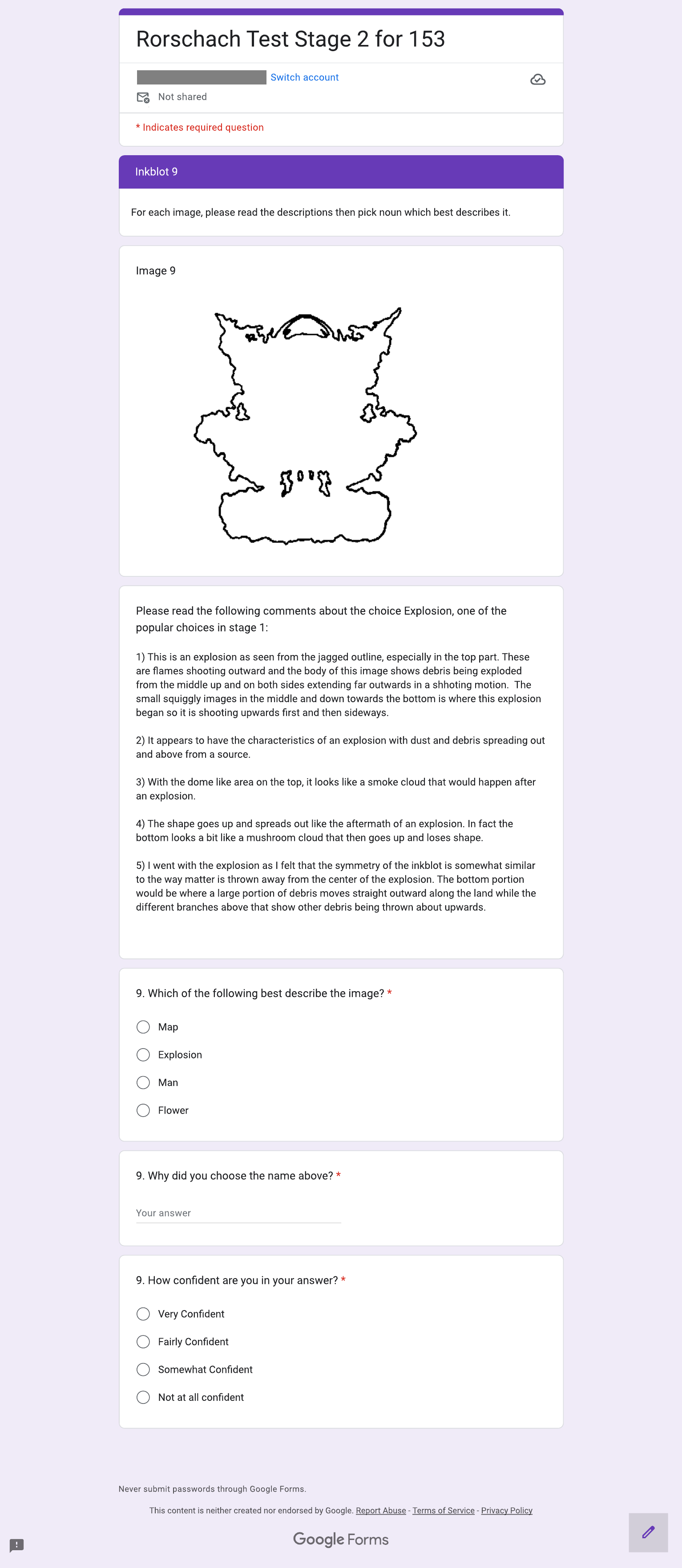}
    \caption{\textbf{Example survey for the CM condition, inkblot 9} Here we show what a survey would have looked like for a participant of the CM condition in our study. While only one is shown for brevity, each inkblot had its own page with the same information presented.}
    \label{sm_fig:example_survey_CM_condition}
\end{figure}

\begin{figure}
    \centering
        \includegraphics[height=.85\textheight]{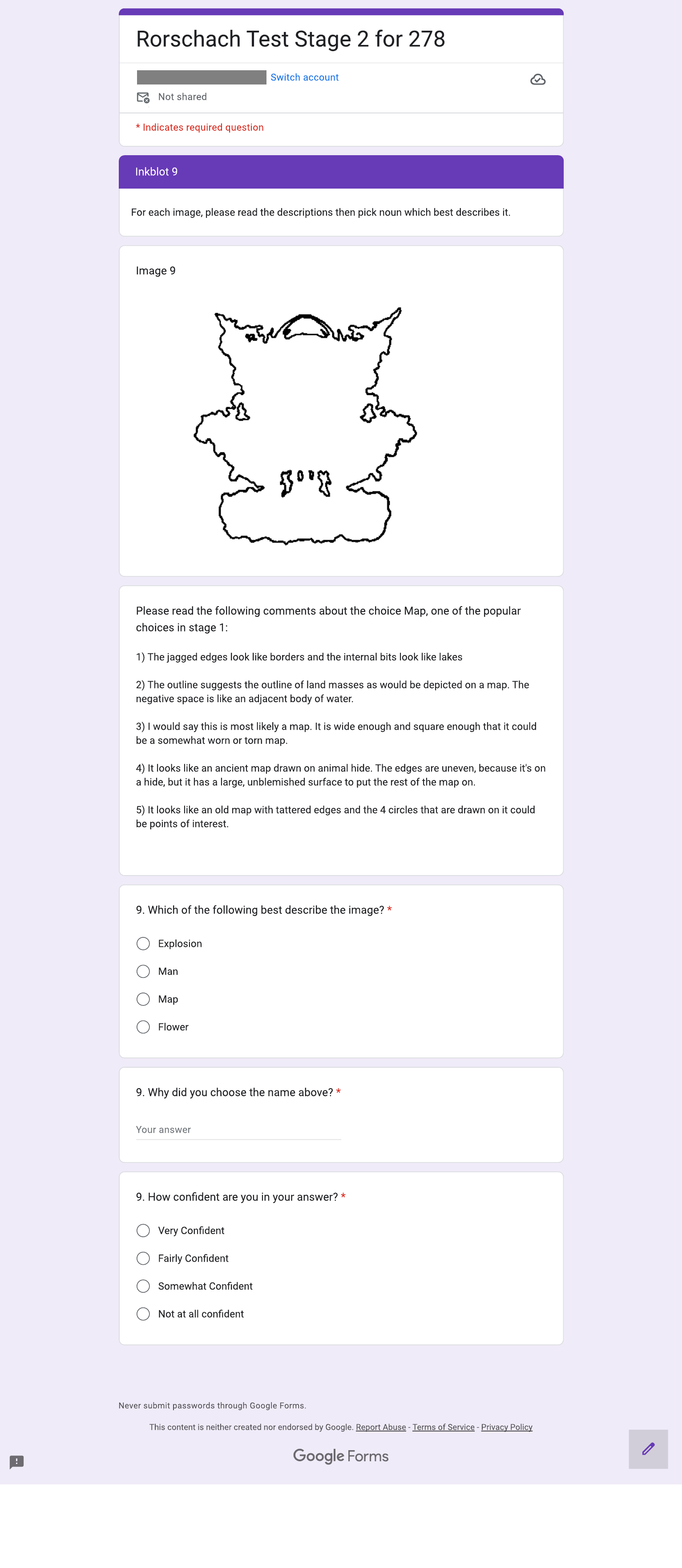}
    \caption{\textbf{Example survey for the CS condition, inkblot 9} Here we show what a survey would have looked like for a participant of the CS condition in our study. While only one is shown for brevity, each inkblot had its own page with the same information presented.}
    \label{sm_fig:example_survey_CS_condition}
\end{figure}

\pagebreak

\section{Tables}

\begin{table}[hbp!]
\resizebox{1\textwidth}{!}{
\begin{tabular}{ll|l|l|lll|lll}
\multicolumn{1}{l}{}          &   & Control                & Recall                 & \multicolumn{3}{c|}{Most-Popular}                                                           & \multicolumn{3}{c}{Second-Most-Popular}                                                    \\
\multicolumn{1}{l}{}          & Opinion
& Total & Total & Total & Promoted& Promoted/Total& Total & Promoted& Promoted/Total\\ \hline
 & $O_1^1$
& 0.242 & 0.163 & 0.154 & N/A & N/A & 0.436 & 0.368 & 0.844 \\
& $O_2^1$
& 0.420 & 0.329 & 0.547 & 0.509 & 0.931   & 0.213 & N/A & N/A \\
& $O_3^1$
& 0.568& 0.452& 0.653& 0.505& 0.773& 0.691& 0.423& 0.612\\

\end{tabular}}
\caption{\textbf{Opinion rank transition frequencies by the first phase opinion for each condition}. 
This table shows the frequency that each phase one opinion ($O_i^1)$ switches between the first and the second phase. There is a notable increase in opinion changing as a function of opinion rank present in most conditions, as well as a decrease in transitioning to the promoted opinions of the CM and CS condition, as well as for all of CR (which promotes all opinions). Additionally, for the CM and CS conditions, the promoted and promoted/total columns show what percentage of switches select the promoted opinion in the second phase and what percentage of total movement that constitutes. This amount is substantial, with roughly 80\% of movement going into the promoted opinion of CM and roughly 70\% for CS. This provides further hints at some strong effect of influence on opinion selection.}
    \label{sm_tab:opinion_switching_rates_by_condition}
\end{table}

\begin{table}[hp!]
\centering
\resizebox{1\textwidth}{!}{
\begin{tabular}{c|llll|} \hline
\multicolumn{5}{|c|}{C Opinion Transitions (P1 -\textgreater P2)}\\
\hline
\multicolumn{1}{|l}{}              &       \multicolumn{4}{c|}{\textbf{Phase 2 Rank}}\\ \hline
\multicolumn{1}{|l|}{}              &       & \textbf{1}       & \textbf{2}      & \textbf{3}     \\
\multicolumn{1}{|l|}{\multirow{3}{*}{\textbf{Phase 1 Rank}}}& \textbf{1} & 275     & 50     & 38\\
\multicolumn{1}{|l|}{}                                  & \textbf{2}     & 47      & 84     & 14\\
\multicolumn{1}{|l|}{}                                  & \textbf{3}     & 31& 15& 56\\ \hline
\end{tabular}
\centering
\begin{tabular}{c|llll|} \hline
\multicolumn{5}{|c|}{CR Opinion Transitions (P1 -\textgreater P2)}\\
\hline
\multicolumn{1}{|l}{}              &       \multicolumn{4}{c|}{\textbf{Phase 2 Rank}}\\ \hline
\multicolumn{1}{|l|}{}              &       & \textbf{1}       & \textbf{2}      & \textbf{3}     \\
\multicolumn{1}{|l|}{\multirow{3}{*}{\textbf{Phase 1 Rank}}}& \textbf{1} & 292     & 27     & 30\\
 \multicolumn{1}{|l|}{}                                 & \textbf{2}     & 40      & 112     & 15\\
 \multicolumn{1}{|l|}{}                                 & \textbf{3}     & 25& 8& 51\\ \hline
\end{tabular}}
\newline
\centering
\resizebox{1\textwidth}{!}{
\begin{tabular}{c|llll|}
\hline
\multicolumn{5}{|c|}{CM Transitions (P1 -\textgreater P2)}\\
\hline
\multicolumn{1}{|l}{}              &       \multicolumn{4}{c|}{\textbf{Phase 2 Rank}}\\ \hline
\multicolumn{1}{|l|}{}              &       & \textbf{1}       & \textbf{2}      & \textbf{3}     \\
\multicolumn{1}{|l|}{\multirow{3}{*}{\textbf{Phase 1 Rank}}}& \textbf{1} & 296     & 30     & 24\\
 \multicolumn{1}{|l|}{}                                 & \textbf{2}     & 81      & 72     & 6\\
 \multicolumn{1}{|l|}{}                                 & \textbf{3}     & 51& 9& 41\\ \hline
\end{tabular}
\centering
\begin{tabular}{c|llll|} \hline
\multicolumn{5}{|c|}{CS Opinion Transitions (P1 -\textgreater P2)}\\
\hline
\multicolumn{1}{|l}{}&  \multicolumn{4}{c|}{\textbf{Phase 2 Rank}}\\ \hline
\multicolumn{1}{|l|}{}              &       & \textbf{1}       & \textbf{2}      & \textbf{3}     \\
\multicolumn{1}{|l|}{\multirow{3}{*}{\textbf{Phase 1 Rank}}}& \textbf{1} & 199     & 130     & 24\\
\multicolumn{1}{|l|}{}                                  & \textbf{2}     & 21      & 118     & 11\\
\multicolumn{1}{|l|}{}                                  & \textbf{3}     & 21& 41& 35\\ \hline
\end{tabular}}
\caption{\textbf{Opinion transition counts}. These tables shows the opinion transitions 
$O_i^1 \to O_j^2$ for some opinion i in the first phase to some opinion j in the second phase. Each table shows this transition matrix for one of the conditions.}
    \label{sm_tab:opinion_transition_counts}
\end{table}